%% file: ml.tex
\newif\ifmulticol	\multicolfalse
\newif\ifshowgit	\showgitfalse		% switches footer on/off
\newif\ifgitlocal	\gitlocalfalse		% use local file gitHeadLocal.gin
\newif\ifbiblatex	\biblatextrue		% defaults to bibtex if false
\newif\ifbibnum		\bibnumtrue 		% num => superscripts, otherwise auth date
\newif\ifbibsort	\bibsortfalse		% biblatex num sort in order of occurrence
\newif\iflineno		\linenofalse
\newif\iftoc		\tocfalse

\newif\iflucida		\lucidatrue
\newif\ifcm			\cmfalse
\newif\ifcharter	\charterfalse		% use for arXiv, use xelatex
\newif\ifcharterotf	\charterotffalse	% requires xelatex

\newcommand*{\secnumdepth}{0}			% sets secnumdepth in \mymaketitle
\newcommand*{\tocdepth}{2}				% sets tocdepth in \mymaketitle
\newcommand*{\reftitle}{References}		% title for references section
\newcommand*{\mytitleskip}{-0.2in}		% change skip after title

\multicoltrue\showgittrue\gitlocaltrue\biblatexfalse\bibnumtrue\lucidafalse\chartertrue

\newcommand*{\mydocfontsize}{\ifcharter11pt\else10pt\fi}
\newcommand*{\setcol}{\ifmulticol twocolumn\else onecolumn\fi}

\documentclass[\mydocfontsize,\setcol]{article}

\newcommand*{\pdfauthor}{\null}
\newcommand*{\pdftitle}{Circuit design in biology and machine learning. I. Random networks and dimensional reduction}

\usepackage{afterpage}
\usepackage{setspace}
\newcommand{\reviset}[1]{\textcolor{teal}{#1}}
\newenvironment{revisetenv}{\color{teal}}{}
\renewcommand{\reviset}[1]{\textcolor{black}{#1}}
\renewenvironment{revisetenv}{\color{black}}{}

\input ml.sty
\input ml-def

\hypersetup{linkcolor={Maroon4}, urlcolor={Maroon4}}

\newcommand*{\mytitle}{\pdftitle}

\newcommand*{\myauthor}{Steven A.\ Frank}
\newcommand*{\myaddress}{Department of Ecology and Evolutionary Biology, University of California, Irvine, CA 92697--2525, USA}
\newcommand*{\mycontact}{\href{https://stevefrank.org}{https://stevefrank.org}}

\newcommand*{\mykeywords}{Biological design, evolution, artificial intelligence, reservoir networks, recurrent neural networks}
\setabstract{-0.07}{\iftoc-.0\else0.0\fi}{%
A biological circuit is a neural or biochemical cascade, taking inputs and producing outputs. How have biological circuits learned to solve environmental challenges over the history of life? The answer certainly follows Dobzhansky's famous quote that ``nothing in biology makes sense except in the light of evolution.'' But that quote leaves out the mechanistic basis by which natural selection's trial-and-error learning happens, which is exactly what we have to understand. How does the learning process that designs biological circuits actually work? How much insight can we gain about the form and function of biological circuits by studying the processes that have made those circuits? Because life's circuits must often solve the same problems as those faced by machine learning, such as environmental tracking, homeostatic control, dimensional reduction, or classification, we can begin by considering how machine learning designs computational circuits to solve problems. We can then ask: How much insight do those computational circuits provide about the design of biological circuits? How much does biology differ from computers in the particular circuit designs that it uses to solve problems? This article steps through two classic machine learning models to set the foundation for analyzing broad questions about the design of biological circuits. One insight is the surprising power of randomly connected networks. Another is the central role of internal models of the environment embedded within biological circuits, illustrated by a model of dimensional reduction and trend prediction. Overall, many challenges in biology have machine learning analogs, suggesting hypotheses about how biology's circuits are designed.

\vskip14pt

\textbf{Teaser text:} Life's biochemical and neural circuits often face the same problems as machine learning: environmental tracking, homeostatic control, dimensional reduction, and classification. What can computational circuits teach us about the design of biological circuits? One insight is the surprising power of randomly connected networks. Another is the central role of internal models of the environment embedded within circuits. Overall, many challenges in biology have machine learning analogs, suggesting hypotheses about how biology's circuits are designed.
}

\begin{document}

\mymaketitle

%% 1st parm is skip on left column at start of TOC, 2nd param is skip after TOC
\iftoc\mytoc{-24pt}{\newpage}\fi

\section{Introduction}

Biological circuits process information and make decisions. For example, a metabolic response circuit may encode the recent availability of various nutrients into a low-dimensional internal representation within the cell. The circuit then decodes the internal state into an appropriate metabolic response \autocite{wu22cellular}.

Similarly, machine learning circuits also process information and make decisions. Particular computational circuits reduce dimension, classify inputs, predict environments, and respond accordingly \autocite{goodfellow16deep}.

Natural and artificial systems design their circuits by learning. Biology uses natural selection's trial-and-error learning. Machines use various learning algorithms. In each case, credit for performance feeds back to circuit components. Update rules reshape the circuits to improve future performance.

How similar are biology's and machine learning's circuits? Mostly, we do not know. I review analogies and highlight topics for study.

Importantly, what we see in machines suggests what to look for in biology. And what works in biology provides insight into machines. We gain a broader perspective on generalized notions of learning and the circuit architecture of learned designs \autocite{kanwisher23using,aldarondo24a-virtual}.

What about intelligence? Some people have argued that machine learning exploits correlations whereas biological intelligence uses a causal model of the world \autocite{mitchell23aischallenge}. But what if machines can discover and use causal models? In any case, the comparison between machines and biology forces clearer thinking about how learning shapes response circuit design.

\subsection{Induction by comparative example}

The distinct materials of machines and biology inevitably cause differences in speed, efficiency, work, and cost. But, for \reviset{for relatively simple} challenges, I do not know of limitations on information processing that require machines and organisms to solve problems with significantly different circuit designs.

If anything, biology's slow chemistry and evolutionary algorithms for learning seem more constrained than machine learning's fast electricity and diverse algorithms, whereas biology's intrinsic parallelism and energy efficiency pose some engineering challenges for machines. These material differences must cause low-level mechanistic differences in response circuits. But we do not know whether high-level structural and functional differences often arise by necessity \autocite{baltussen24chemical}.

Put another way, we lack \reviset{well-developed theory} that predicts essential differences between the domains. We must \reviset{proceed} inductively by accumulating various comparisons between machine learning and biology. How do the distinct systems solve particular challenges? Architecturally, what sorts of similarities and differences do we see in the design of response circuits? Are there qualitative differences in information flow and decision making?

Ultimately, what sorts of predictions can we make about the design of biological response circuits? What sorts of inductive generalities arise about how learning shapes circuits and systems?

To start on these questions, I begin with some basic machine learning circuits. I discuss possible links between each computational example and biological circuits. I emphasize simple genetic and biochemical circuits in cells, for which there are relatively few prior studies of this sort. I occasionally mention neurobiological networks, for which there are many previous analyses that link computational and biological circuits \autocite{mcculloch43a-logical,hebb49theorganization,rumelhart86parallel,jiahui23modeling,kanwisher23using}.

\subsection{Overview}

Each article in this series develops particular comparisons between machines and biology. This first article starts with randomly connected reservoir networks in machine learning. Such reservoir networks can be remarkably successful for certain types of challenges. For example, a randomly connected circuit can often predict future inputs given the recent sequence of input values. 

Reservoir networks introduce several general principles of circuit design. The following section summarizes those principles. After that, I consider possible biological insights from the study of random networks in machine learning. 

One insight concerns the evolutionary origin of specific circuit designs. Consider perception and response. Perception provides little value without a matching response. How do things get started? Machine learning's random reservoir networks suggest a path.

Random networks also illustrate how circuits gain from reducing the dimensionality of their inputs. The second part of this article continues analysis of dimensional reduction, a process that benefits circuit designs for many different kinds of challenges.

For example, recognizing an ear as a facial component in an image is the reduction of a high-dimension array of pixels to a lower-dimension model of a face. Similarly, a biological cell may reduce the abundances of various nutrients to a single dimensional measure of beneficial metabolic state. Machine learning models provide many insights into how simple circuit designs can reduce dimension to solve challenges.

The next article in this series discusses anomaly detection. In machine learning, a particular challenge might be detecting a rare fraudulent credit card transaction among a vast number of legitimate transactions. What pattern of inputs suggests that something is atypical? The problem is easy if certain features associate with danger. The problem is hard if anomalies often arise in different ways so that all one can do is identify anomalies by an atypical multivariate pattern.

A similar challenge arises in biology. Many particular signs of danger can be detected relatively easily. But how does an organism identify trouble by sensing a nonspecific deviation from the typical multivariate pattern of inputs? The methods and examples from machine learning provide general insights into the circuit designs that can solve this challenge.

For example, in machine learning, what characteristics and patterns of sensor arrays help in detecting anomalies? How does combining information from different circuits in an ensemble improve success? What is the role of sequential improvement in circuit performance, an analogy with the evolutionary dynamics of circuit improvement in biology? What hypotheses do we gain about how simple biological circuits may be designed?

Future articles in the series will explore such questions. Topics include how circuits manage uncertainty, how they learn to embed models of the world, and how machine learning techniques aid in designing simple biochemical circuits. Another topic examines how robustness influences circuit design and complexity.

\section{Reservoir computing}

\begin{figure*}[t]
\centering
\hbox{\null}\vskip-24pt
\includegraphics[width=\hsize]{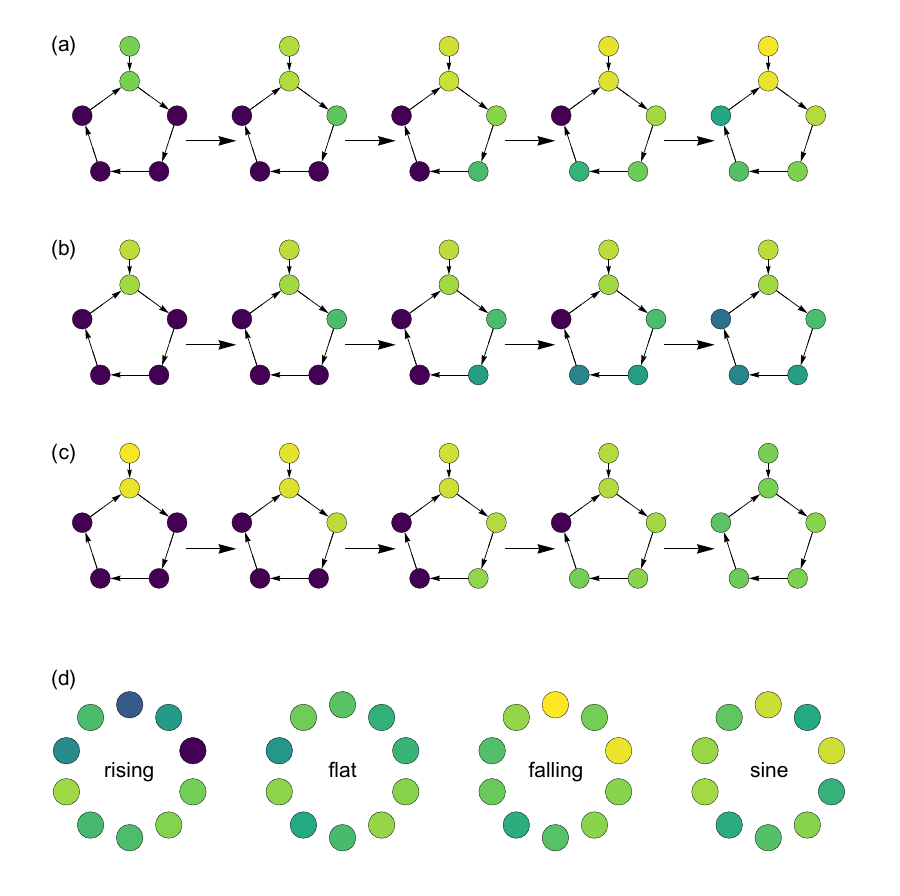}
\vskip0pt
\caption{\reviset{Reservoir networks provide information about past inputs. The simple examples here illustrate the process. The intensity of shading for each node reflects its value, with lower values having darker shading and darker color. (a) The five nodes in this example create a circular network \autocite{rodan10minimum}. In each timestep, each node passes its value, $v$, transformed by $\tanh(v/2)$, to the next node, in which $\tanh$ is the hyperbolic tangent function. That function maps its input to the range $[-1,1]$, keeping the values in the network bounded to a useful set. The node at the top provides the external input into the network, with input value transformed by $\tanh$. In this example, the inputs $(0.1,0.3,0.5,0.7,0.9)$ rise linearly over time. As the inputs proceed from left to right, the network values encode information about the input sequence. The final state on the right has declining values around the circle when starting from the top node, associated with the rising inputs over time. (b) Network dynamics for flat inputs, each of $0.7$. In the final state, nodal values decline more slowly. (c) Network dynamics for falling inputs, which are the reverse order of the inputs in (a). In the final state, nodal values are relatively flat, balancing the decline of input values against the decline in values transmitted around the circle. (d) A reservoir network with ten bidirectionally connected nodes. Each circle of nodes shows a different instance of the same network. The weights connecting the nodes are random, and the weights connecting each input to each node are also random. The same network connectivity is used for each instance. The input pattern for an instance is described at its center. In this case, each input sequence has 11 values. The illustrations show the final state after all inputs. The conclusion is that a network's final state reflects its input pattern, providing the system with information about the history of inputs.}}
\label{fig:resBack}
\end{figure*}

This section illustrates how randomly connected networks store information about recent inputs. That reservoir of information about the past can be used to predict future environmental states and to respond accordingly. Random reservoir networks first arose in computational modeling \autocite{maass02real-time,jaeger07echo,gauthier21next,cucchi22hands-on}. Several studies then considered how biology exploits similar reservoir networks \autocite{damicelli22brain,goudarzi13dna-reservoir,yahiro18a-reservoir,seoane19evolutionary,czegel21novelty,loeffler21modularity,sole22evolution}. \reviset{Box 1 and \Fig{resBack} provide background on reservoir computing.}

Reservoirs illustrate five important network attributes, described in more detail below. First, reservoirs have feedback connections among nodes, a kind of recurrent neural network. Recurrence provides the capacity for memory, which allows calculations over past input sequences  \autocite{goodfellow16deep}.

Second, a reservoir's random connectivity transforms inputs to a lower dimensional internal representation.

Third, the circuit often functions precisely in spite of the internal randomness in design \autocite{frank23precise}.

Fourth, at a small scale, the individual nodes and their connectivity may be nearly uncorrelated with large-scale circuit function. 

Fifth, large networks may be overwired. They often have more parameters and connectivity than required for the problem being solved \autocite{sorrells15making,frank17puzzles5}.

Machine learning's reservoir model provides analogies for biological circuits. Before turning to those analogies for biology, I summarize additional aspects of reservoir computing.

\subsection{How do random networks store information?}

Think of a liquid with a smooth surface. Drop a stone on the surface to generate waves. Then drop another stone, and another. The pattern of surface waves contains information about the temporal history of inputs.

A reservoir's computational network acts similarly. Each input propagates signals through the recurrent network based on the random connectivity pattern between nodes and the rules for updating nodes. At each point in time, the network state contains information about the history of inputs \autocite{maass02real-time}.

\subsection{Prediction and response}

In a computational reservoir model, the circuit can be trained to predict the future values of environmental variables. The randomly connected network takes in a sequence of inputs. The internal states of the network's nodes reflect the past input sequence. The machine learning algorithm then fits a regression model, using the reservoir's current state at any point in time to predict the future values of the environmental variables \autocite{chattopadhyay20data-driven,frank23precise,chepuri24hybridizing}.

\begin{figure*}[t]
\centering
\includegraphics[width=0.75\hsize]{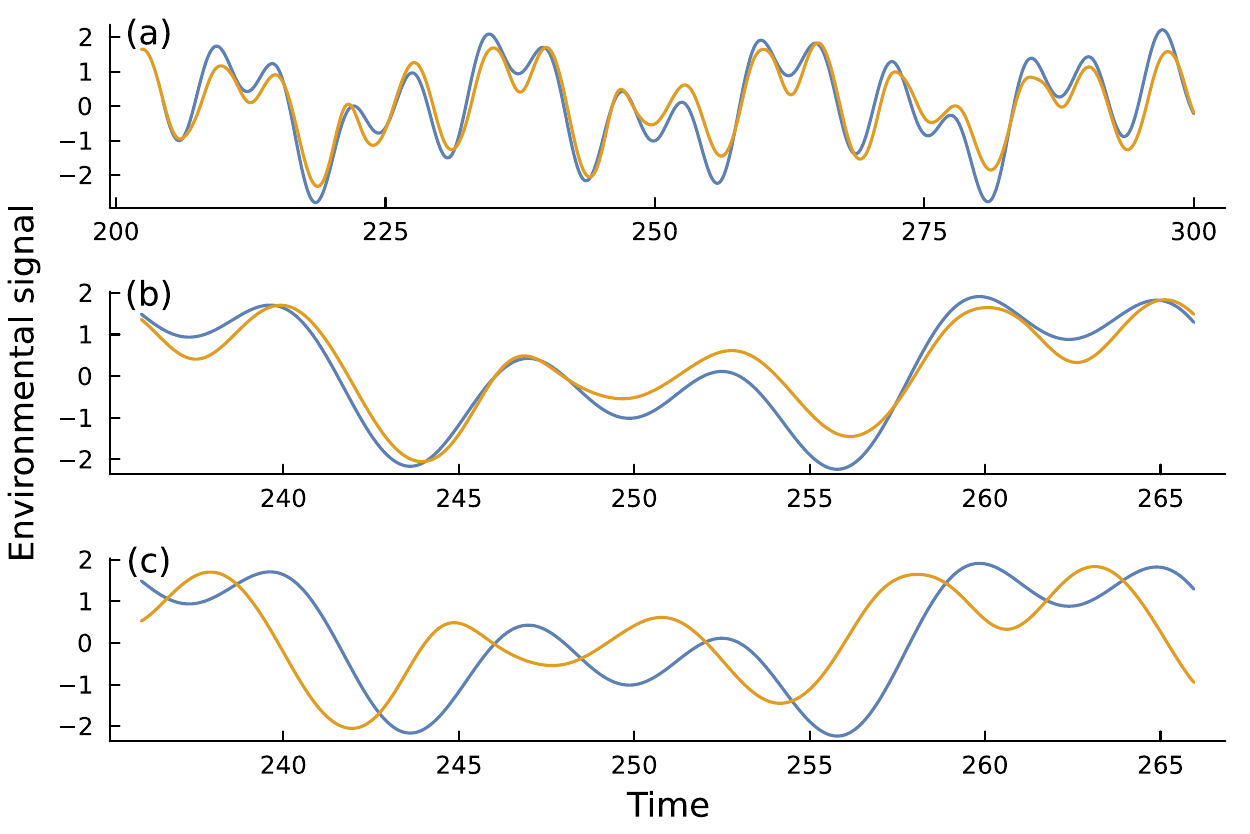}
\vskip0pt
\caption{Match between actual (blue) environmental inputs and predicted (gold) inputs generated by a reservoir network. See text for description of each plot. Briefly, (a) the match over 100 nondimensional time units, (b) magnification of a short time sequence, and (c) gold curve shifted 2 units to the left.  I created the reservoir networks with the Julia programming language package ReservoirComputing.jl \autocite{martinuzzi22reservoircomputing.jl:}. Each network has 20 nodes. The connectivity matrix was randomly generated with a $0.6$ expected frequency of zeros. The connectivity weights were normalized to a matrix spectral radius of $1.0$, which means that the nodal values tend neither to increase nor decrease over time. The input is $u(t) = \sin(t)+\sin(0.51t)+\sin(0.22t)$. The total input sequence occurred over time units $t=[1,300]$ in increments of $\GD t=0.05$. In the training period during the first 200 time units, I fit a regression model on the nodal values, $\bmr{x}$ in \Eq{rnn}, to predict the input values at 2 time units into the future, which is $2/\GD t=40$ sequence increments. To fit the prediction model, I used lasso regression with an L1 regularization cost of $0.0003$. That linear cost on the magnitude of the parameters prevents large parameter values and reduces to zero those parameters that contribute little to the predicted fit. In this case, the fitted regression used 7 of the 20 nodal values, a significant reduction in dimensionality of the input complexity. Reducing the L1 cost to zero used all 20 nodal values in the fitted regression and gave a nearly perfect fit. The plots show the results for the inputs that were not used during the fitting procedure. The freely available Julia computer code provides full details about assumptions and methods for all figures in this article \autocite{frank24circuit-code}.}
\label{fig:reservoir}
\end{figure*}

 \Figure{reservoir} illustrates the response of a reservoir network. The blue curve in (a) traces an input signal. The gold curve shows the network's prediction for input values. Panel (b) magnifies part of panel (a). 

Panel (c) clarifies the nature of predictions. At each point in time, the blue curve marks the input value. The gold curve shows the predicted input 2 time units into the future. Panels (a) and (b) shift the gold curve 2 time units to the right so that deviations between the curves measure the difference between the actual and predicted future inputs. Time units are nondimensional and may be interpreted on whatever scale makes sense for a particular application.

Here are the details \autocite{martinuzzi22reservoircomputing.jl:}. A simple reservoir network has $N$ internal nodes, with nodal values $\bmr{x}_t$ at time $t$. A random matrix, $\bmr{W}$, describes the connection weights from each node to all nodes. The $m$ input values at each time are $\bmr{u}_t$. The matrix $\bmr{W}_{\bmr{in}}$ describes the connection weight from each input to each network node. The update rule for nodal values is
\begin{equation}\label{eq:rnn}
  \bmr{x}_{t+\GD t}=(1-\Ga) \bmr{x}_t + \Ga\mskip1mu\tanh\lr{\bmr{W}\bmr{x}_t 
  		+ \bmr{W}_{\bmr{in}}\bmr{u}_{t+\GD t}}.
\end{equation}
The function $\tanh$ returns the hyperbolic tangent of its argument. The parameter $\Ga$ sets the rate at which nodal values update. In reservoir models, all parameters remain constant. In later models based on this equation, the learning phase adjusts these parameters.

For simple reservoir computing, one fits a regression model that maps the nodal values at a particular time to a future input value. Training a network to make predictions follows standard machine learning protocol. Split a given sequence of inputs into a training part and a test part. Use the training part to fit the model parameters. Use the test part to analyze how well the system performs on data not used for training. See the figure caption for more detail about this particular case.

This example shows that a simple randomly constructed reservoir can store sufficient information about past inputs to make good predictions about future inputs. Reservoir networks have solved a wide variety of challenging problems \autocite{chattopadhyay20data-driven,frank23precise,chepuri24hybridizing,cucchi22hands-on}. \reviset{For example, such networks predict macroeconomic variables \autocite{ballarin24reservoir} and spatiotemporal patterns of sea surface temperatures \autocite{walleshauser22predicting} and solar irradiance \autocite{li20multi-reservoir}. Other studies suggest that variants of reservoir computing networks provide good models for neurobiological circuits \autocite{loeffler20topological,loeffler21modularity}.}

\section{Biological insights from reservoirs}

\subsection{Recurrence, memory, state, and dimension}

Reservoirs have four attributes that make them particularly good at using past inputs to predict future inputs. Those four attributes are likely to shape the architecture of biological circuits.

\textit{Recurrence} means that information flows through feedback loops within a network.
In machine learning, one thinks of a network as composed of connected nodes, each node contributing calculations to the network's overall computational path.

In the simplest feed forward architecture, information flows unidirectionally, starting from the external inputs and flowing into the first layer of computational nodes. The outputs of that first layer feed forward as inputs to the next layer, and so on, until the final nodal outputs form the circuit's response.

If a node feeds its output back as input to itself, that is a recurrent feedback loop. Such a node gets inputs from different time traces of information flow. A feed forward input represents a relatively current time trace. The self-feedback represents a prior time trace. In general, recurrence occurs when a node receives input from multiple time traces, with feedback loops bringing in earlier time traces.

Recurrence creates a kind of \textit{memory}. With a recurrent loop, the input flowing into a particular node may include both the first trace of information from the most recent external input into the network and traces of information from earlier inputs into the network. The temporal offset created by recurrent flow allows a node to compute on sequences of inputs.

At the network level, memory can arise simply by recurrent connectivity. Each computational node takes inputs and produces outputs without any local memory. Alternatively, nodes may retain memory through their internal \textit{state}. For example, a node that retains memory of its most recent output can use that internal state as input for its next computation. Nodes in reservoir networks may retain internal state, as in \Eq{rnn}.

\reviset{Thus, we have two distinct types of memory. Recurrent network memory is a global property of the network, arising from feedback connections that create the possibility for the inputs into a node ultimately to trace back to external environmental inputs that came into the network at different times. By contrast, local state memory is an internal property of a single node by which the node retains information about past inputs into that node.}

Reservoirs perform \textit{dimensional reduction} of their inputs. The temporal input sequence may be a high-dimensional vector. The reservoir's internal nodal states reduce that high-dimensional input to a lower dimension.

Because computational reservoirs are typically created with random, unchanging connectivity, many of the internal states may either be redundant or not useful with regard to the particular computational challenge. In reservoir computing, the overall computation picks out the useful states to make a prediction about the challenge, often using regression on the states to predict an outcome. The prediction stage typically reduces the internal state dimensionality to a much lower dimension of useful information.

Recurrence, memory, state, and dimension provide a basis for evaluating biological network architecture.

\subsection{Form and function, discovery, and refinement}

Computational reservoirs raise three interesting questions about biology. 

First, how does a circuit's form relate to its function? The fact that simple random reservoirs perform well suggests that, for circuits, the low-level connectivity patterns that determine form may often be weakly correlated with function. Good performance arises from the nonspecific attributes of recurrence, memory, state, and dimensional reduction, which transcend the specific details of connectivity between nodes.

Second, how do things get started? Natural selection can only refine the performance of something that already exists and has the capacity for minor adjustments to improve function. Random reservoirs may be a first step in creating functional circuits, providing the basis for something that initially works and can subsequently be improved.

Third, if initial discovery arises by a network that is randomly connected, how effective is natural selection at subsequently refining the circuit architecture? It may be that random connectivity happens only in the initial phases of a circuit's evolution. Subsequently, selection may refine the circuit, hiding the initial randomness from view. Or the initial random connectivity may be sufficiently good, making further selection too weak to modify form.

Overall, how often do circuits first arise by co-opting an existing network that is initially random in its connectivity with respect to function? How often will subsequent selection erase traces of that initial randomness? In other words, what sorts of relations between form and function should we expect at different stages in a circuit's evolution? The following examples and comments suggest how we might start on these questions.

\subsection{Origin of traits: perception and response}

The origin of perception-response traits poses a puzzle. Perception provides no benefit without a response. A response cannot happen without perception. How do things get started? That question guides how we may think about form and function, discovery, and the refinement of biological circuits.

Reservoir computing suggests a solution. If, initially, a randomly connected network provides information about environmental inputs, then a biological circuit can subsequently use that perceived information to evolve an appropriate response \autocite{frank23precise}. Once a working perception-response circuit is in place, natural selection may refine the circuit.

The prior section showed how a random reservoir might initiate a new perception-response circuit. In \Fig{reservoir}, the system takes in a single input that fluctuates in a complexly periodic way over time. A sparsely and randomly connected network with 20 nodes stores an imperfect and dimensionally reduced memory of past inputs. That memory provides a sufficient basis for a population of organisms to learn, by natural selection, how to predict future environmental states using only the internal states of its randomly connected network. Many studies show that random networks provide the foundation for predicting complex patterns of future environmental fluctuations \autocite{loeffler20topological,loeffler21modularity,chattopadhyay20data-driven,frank23precise,chepuri24hybridizing,cucchi22hands-on}.

Given that random networks often perform very well, how much might we expect natural selection to refine such circuits? There is a tendency in the systems biology literature to expect that natural selection refines every small-scale feature of a biological circuit. However, at a small scale, the intensity of natural selection may often be rather weak relative to various stochastic evolutionary processes. Circuits may often be more randomly or more complexly wired than expected from a purely optimizing learning process \autocite{munoz-gomez21constructive,frank23robustnessb,lynch24evolutionary}.

\subsection{Prokaryotes, eukaryotes, and multicellularity}

When a novel challenge arises, organisms may benefit from a pre-existing network that is randomly connected with respect to the new challenge. What might influence the chance that there is a usable pre-existing network?

Comparing prokaryotes and eukaryotes provides an interesting contrast. Prokaryotes typically have smaller cells that are more limited in their total protein content. Proteome limitation may constrain the transcription factor network that comprises the primary cellular response circuits \autocite{chen21proteome,frank22microbial}. In addition, populations of prokaryotes may typically be larger than populations of eukaryotes. Larger populations mean that natural selection is more likely to outweigh the random sampling processes of small populations, more effectively refining the details of functional circuits and reducing the amount of overconnected random wiring \autocite{lynch07the-origins,frank23robustnessb}.

Prokaryotic transcription factor networks may therefore be less widely connected and more finely pruned than eukaryotic networks. If so, then prokaryotes may be less likely to have pre-existing networks that are essentially random with respect to new challenges. And prokaryotes may be less likely than eukaryotes to retain random or excessively wired networks after the refining action of natural selection.

Comparing multicellular and unicellular eukaryotes, two factors favor broader and more randomly connected networks in multicellular organisms. First, multicellular populations tend to be smaller, with more random fluctuations that weaken the relative efficacy of natural selection \autocite{lynch07the-origins}. Second, in multicellular organisms, individual cells are tuned for the differentiation of functions rather than the unicellular tuning for fast growth and efficient resource usage. The greater emphasis on speed and efficiency in unicells likely prunes the size and randomness of their circuits.

Overall, broader circuit connectivity and more randomness likely occur in multicellular eukaryotes, followed by unicellular eukaryotes, and then unicellular prokaryotes. Perhaps the diversity and complexity of multicellular eukaryotic cellular circuits provides greater opportunity for the discovery of novel functions.

\subsection{Precise traits from sloppy components}

A trait may first arise from a circuit with reservoir-like random connectivity. Subsequently, natural selection may not be sufficiently strong to refine and prune the network. A circuit with reservoir-like random connectivity appears rather sloppy in its low-level design yet can function precisely with respect to its environmental challenge.

In other words, we may see precise traits arising from sloppy components \autocite{frank23robustnessb}. Other evolutionary pathways can also lead to low-level sloppiness of circuits that nonetheless function in a relatively precise way  \autocite{frank07maladaptation,frank23robustnessb,lynch12evolutionary,lynch24evolutionary,gray10irremediable,daniels08sloppiness,gutenkunst07universally}.

\subsection{Network architecture}

Random connectivity can be effective for certain challenges. However, within reservoir computing, imposing some structure to connectivity patterns often improves circuit performance or reduces the size of the required reservoir \autocite{rodan10minimum,rodan12simple,loeffler20topological,loeffler21modularity}. 

In biological circuits, hierarchy and modularity may occur within pre-existing circuits. Or such broad-scale structure may evolve to refine circuit performance \autocite{tripathi23minimal}. For example, \textit{Caenorhabditis elegans} organizes its 302 neurons into a modular architecture, which may enhance the computational efficacy of its neural circuits \autocite{ruach23thesynaptic}.

\subsection{Neurobiology}

Machine learning models have long been linked to neurobiological circuits \autocite{mcculloch43a-logical,hebb49theorganization,rumelhart86parallel}. Recent articles continue to develop those connections \autocite{jiahui23modeling,kanwisher23using}, including discussion of reservoir computing as a model for particular aspects of neurobiology \autocite{ramezanian-panahi22generative}. Although this article emphasizes genetic and biochemical circuits of cells, it is helpful to mention a few applications of reservoir computing concepts to neural circuits.

For small neural circuits, the connectivity of reservoir wiring may be relatively repeatable from one individual to another. With developmentally predictable wiring, the readout of the reservoir states can also evolve to be fixed developmentally. For larger circuits, the reservoir wiring may vary developmentally between individuals. The internal reservoir states no longer express predictable information about inputs to the circuit. 

In the case of larger circuits with less consistent wiring, the organism must learn to read the circuit's states in relation to the particular environmental problem that the circuit must solve. Learning to read a reservoir circuit may be the origin of commonly observed critical learning periods \autocite{frank23precise}.

At a more abstract level, \textcite{czegel21novelty} ask how complex intelligence may arise from underlying neural circuits. They studied a computational model of initially random reservoir circuits within a single brain that, in effect, teach one another. The ensemble of circuits uses Darwinian trial-and-error selection to learn the solutions to complex challenges.

\section{Dimensional reduction}

Systems often encode high-dimensional inputs to a low-dimensional representation, highlighting important attributes \autocite{reddy20analysis,sorzano14asurvey}. Systems may also contain a model of the environment, which helps to transform an internal representation of inputs into an appropriate action \autocite{scholkopf21toward,bin22internal,gupta23theinternal}. \reviset{Box 2 provides general background for this topic.}

\subsection{Encoder network}

An encoder network in machine learning typically processes a high-dimensional input within its initial layer. That first layer then passes a lower-dimensional output to the next layer of the network. A series of dimension-reducing layers may follow, each layer producing fewer outputs than the number of inputs into the layer \autocite{chen23auto-encoders,mao24thetraining}.

By forcing the system to reduce dimensionality through a series of partial steps, each reducing layer has the opportunity to capture different essential features of the initial input. The sequential reduction sometimes finds efficient and effective representations that can be difficult to find when reduction is attempted in a single step.

The input may occur as a temporal sequence. If so, then the initial network layer stores a memory of past inputs in order to find an effective reduction of the input sequence. A memory trace can be created by recurrent network connections, like those described by \Eq{rnn}.

\subsection{Predicting direction of change}

Suppose, for example, that a system experiences a sequence of environmental inputs. The system gains by predicting the direction of change for the next input relative to the current environment. Prior inputs provide information on the next direction of change.

I develop an example for which we can obtain the optimal dimensional reduction and internal decision model. We can then observe how a machine learning model discovers and encodes a nearly optimal solution.

For many realistic problems we may not be able to solve for or guess what sort of circuit solution is required. In those cases, machine learning guides us in the design of circuit solutions and the analysis of how biological circuits solve environmental challenges.

Looking out toward those more difficult problems, the basic example developed here illustrates three points. First, very simple networks can learn to encode efficient computational circuits based solely on feedback about current performance. Second, dimensional reduction is a common attribute of effective circuits. Third, circuits may encode a causal model rather than exploit a database of correlations for past inputs. This example extends a previous study \autocite{frank24a-biological}.

\subsection{Optimal solution}

To generate input sequences, start with a random walk, $\hat{u}_t$. Then replace each value at time $t$ by its exponential moving average, $u_t = \Gb \hat{u}_t + \lr{1-\Gb} u_{t-1}$. \Figure{enc_dyn}a,e show two example sequences. The figure legend describes details.

The goal is to predict the direction of change, the sign of $\GD u_t = u_{t+1}-u_t$. The sign is positive when
\begin{equation*}
  \GD u_t = \Gb\GD\hat{u}_t + \lr{1-\Gb} \GD u_{t-1} > 0.
\end{equation*}
Because $\hat{u}_t$ is a random walk, its change $\GD\hat{u}_t$ is equally likely to be positive or negative. Thus, the best prediction for the sign of $\GD u_t$ is the sign of $\GD u_{t-1}$. In other words, the most recently observed direction of change provides the best prediction for the next direction of change.

Roughly speaking, we may think of the recently observed change, $\GD u_{t-1}$, as the trend momentum. Positive momentum means that the trend is likely to continue up. Negative momentum means that the trend is likely to continue down.

The greater the momentum, the more likely the trend will continue in the same direction. For example, the greater $\GD u_{t-1}$, the more negative the underlying random walk change, $\GD\hat{u}_t$, must be to reverse the trend. Increasingly extreme moves in the underlying random walk are increasingly uncommon.

Thus, an ideal internal model uses the currently observed trend direction to predict the next direction of change. And it uses the trend momentum to estimate the confidence in the directional prediction.

\subsection{Circuit that predicts directional change}

A computational circuit can learn to solve this directional prediction problem. The initial layer of the circuit has the recurrent memory structure in \Eq{rnn} with a pool of $N=2^n$ internal states.

For $n\ge2$, I used a sequence of layers with $2^n,2^{n-1},\dots$ for dimensional reduction. In some cases, larger networks learned more quickly. But an initial layer with $2$ internal memory states was sufficient. The circuit needs information only for the current and prior inputs to generate an estimate for the momentum, $\GD u_{t-1}$.

I focused on this smallest 2-state circuit, as shown in \Fig{encoder_net}. The figure legend describes how the circuit transforms inputs into two internal memory states, reduces those memory states to an internal one-dimensional summary value, and then transforms the reduced summary value into a prediction about future change.

\begin{figure*}[t]
\centering
\includegraphics[width=0.98\hsize]{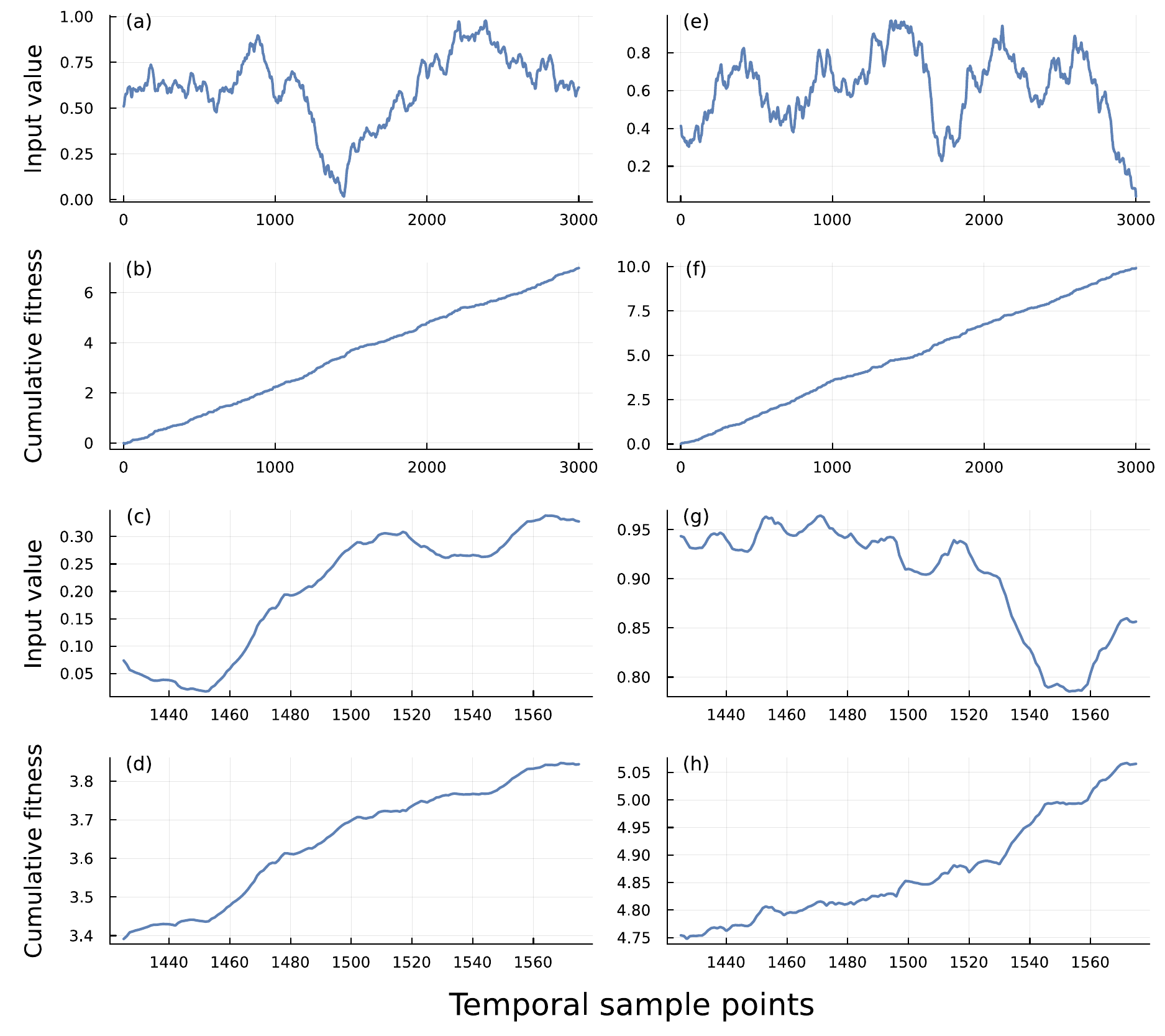}
\vskip0pt
\caption{Encoder circuit prediction for the direction of change in a sequence of observations. The input sequence is calculated by starting with a random walk, $\dd \tilde{u} = \Gs\dd W$, at $\tilde{u}_0=0$ with $\Gs=0.2$, in which $\dd W$ is a Wiener process that samples a normal random variable with standard deviation of $\sqrt{\dd t}$, with $\dd t=0.01$ in all examples. The sequence values are normalized to $[0,1]$ by affine transformation, yielding $\hat{u}$. The sequence is then replaced by its exponential moving average, $u_t = \Gb \hat{u}_t + \large(1-\Gb\large) u_{t-\GD t}$, sampled at discrete time points, and with $u_0=\hat{u}_0$ and $\Gb=0.2$. \Figure{encoder_net} shows the circuit architecture. The 12 network parameters in the \Fig{encoder_net} circuit were adjusted to reduce the loss function by the Adam learning algorithm with learning rate $0.005$ applied to 25,000 randomly generated sequences. The examples in this section analyze 300 time units with 10 sample points per time unit for a total of 3000 sample points. After optimization, new sequences were used to test performance, as follows. (a) Input sequence. (b) The fitness for each prediction is the absolute value of the difference between the current and prior observation multiplied by $-1$ if the prediction about direction is incorrect. Cumulative fitness is proportional to the sum over all predictions up to the current time point. (c,d) A magnified view of a time interval from the plots above. (e--h) Similar plots for a second input sample.
}
\label{fig:enc_dyn}
\end{figure*}

\begin{figure*}[t]
\centering
\includegraphics[width=0.8\hsize]{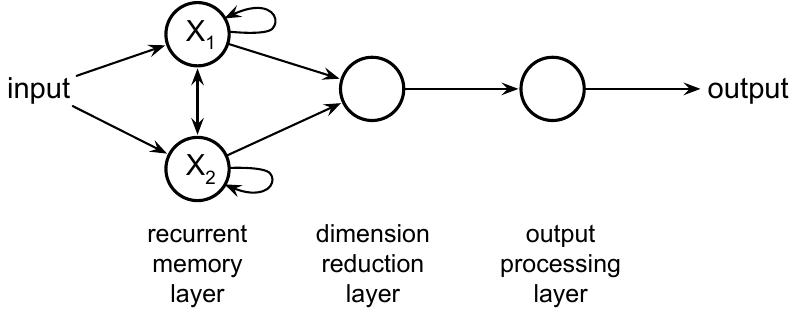}
\vskip7pt
\caption{Encoder dimensional reduction network for predicting sequence trend. In this example, the input is a sequence of values. The recurrent memory layer is described by \Eq{rnn}, allowing the network to calculate information about current and past inputs and retain that information in the internal states, $x_i$. Output from this layer is then passed as input to the dimension reduction layer. That layer sums an affine transformation of each input to produce a one-dimensional output passed to the final layer. The final layer applies to its input an affine transformation followed by a sigmoid function, transforming inputs on $(-\infty,\infty)$ to the final output on $[0,1]$. The output, $\Gr$, is taken as a prediction of the direction of change of the next input value relative to the current input value. An actual positive change in the data associates to a target of $\Gk=1$, and a negative change associates to a target of $\Gk=0$. The distance between the prediction, $\Gr$, and its target, $\Gk$, is the cross-entropy loss function, $-\Gk\log\Gr-(1-\Gk)\log(1-\Gr)$. \Figure{enc_dyn} shows an application of this circuit architecture to a particular example.
}
\label{fig:encoder_net}
\end{figure*}

\Figure{enc_dyn} shows results for one particular circuit. That circuit learned to predict the direction of change by adjusting its 12 internal parameters. The adjustment happened by a standard machine learning cycle. In each cycle, I generated a sample input such as the one shown in \Fig{enc_dyn}a. I then calculated the performance by using the loss function described in the figure legend.

Initially, a circuit predicts the direction of change with a success rate of $0.5$, equivalent to flipping a coin. The computer code then calculates a gradient for how changes in the parameters alter the performance. The algorithm adjusts the parameters to improve performance slightly. The process repeats with a new random input sequence. In this example, I used 25,000 rounds of parameter refinement.

\Figure{enc_dyn}b,f shows the performance of the circuit after the final parameter refinement, measured as cumulative fitness, as described in the figure legend. An upward trend in cumulative fitness associates with an accuracy of prediction for the direction of future change that is greater than the random baseline of $50\%$.

\begin{figure*}[t]
\centering
\includegraphics[width=0.75\hsize]{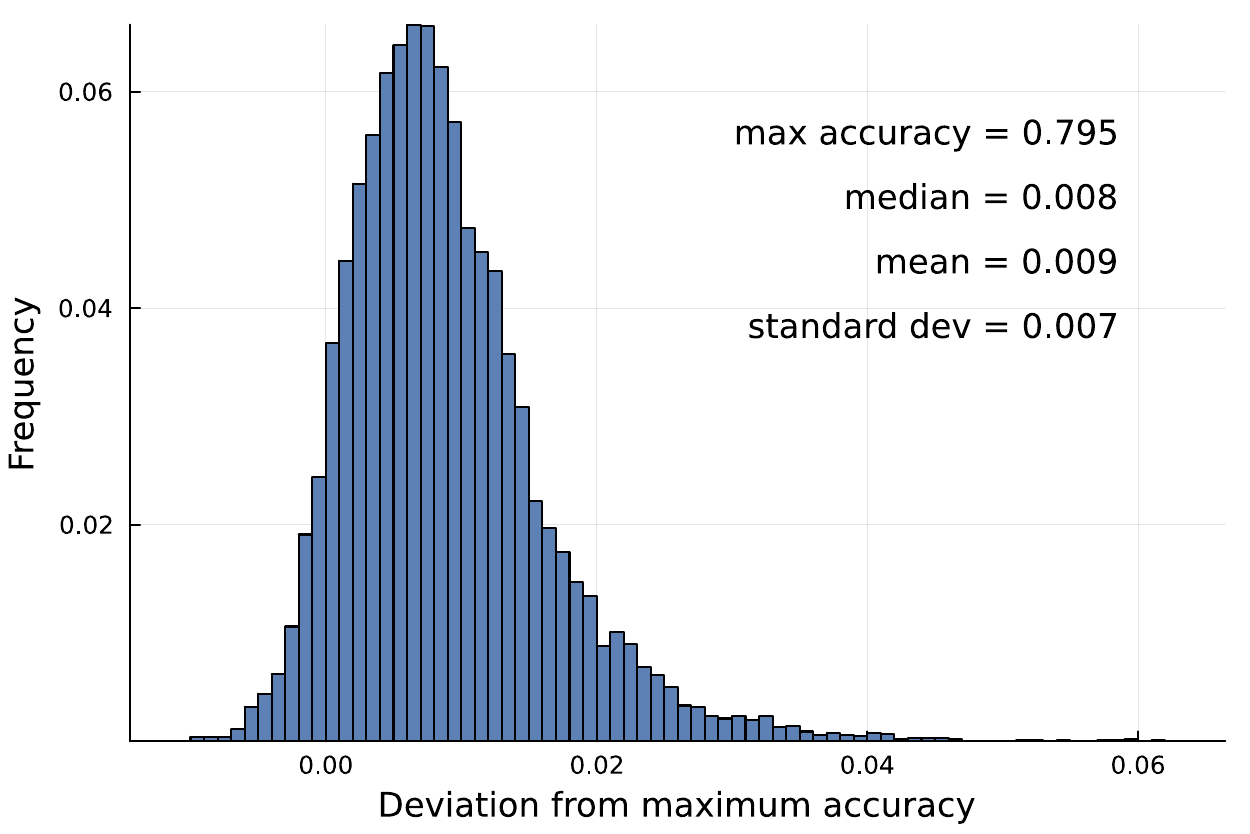}
\vskip0pt
\caption{Circuit accuracy in predicting the direction of input change. Random input sequences were generated as described in \Fig{enc_dyn}. For a particular generated sequence with $100,000$ sample points, the direction of change between two inputs predicted the direction of change for the next input with frequency $0.795$, which estimates the maximum accuracy that could be achieved on that sequence. For each of $10,000$ novel sequences with $3000$ sample points, I calculated the deviation between the optimized circuit's frequency of correct predictions and the maximum estimated accuracy. The histogram shows the distribution of those deviations. The median, mean, and standard deviation refer to that distribution of deviations.
}
\label{fig:enc_err_distn}
\end{figure*}

The method that generated the random sequences creates inputs for which there is an approximately $80\%$ match between the direction of change at one time and the direction of change at the next time. \Figure{enc_err_distn} shows that the circuit correctly predicted the direction of change at nearly the optimal match frequency.

\begin{figure*}[t]
\centering
\includegraphics[width=0.75\hsize]{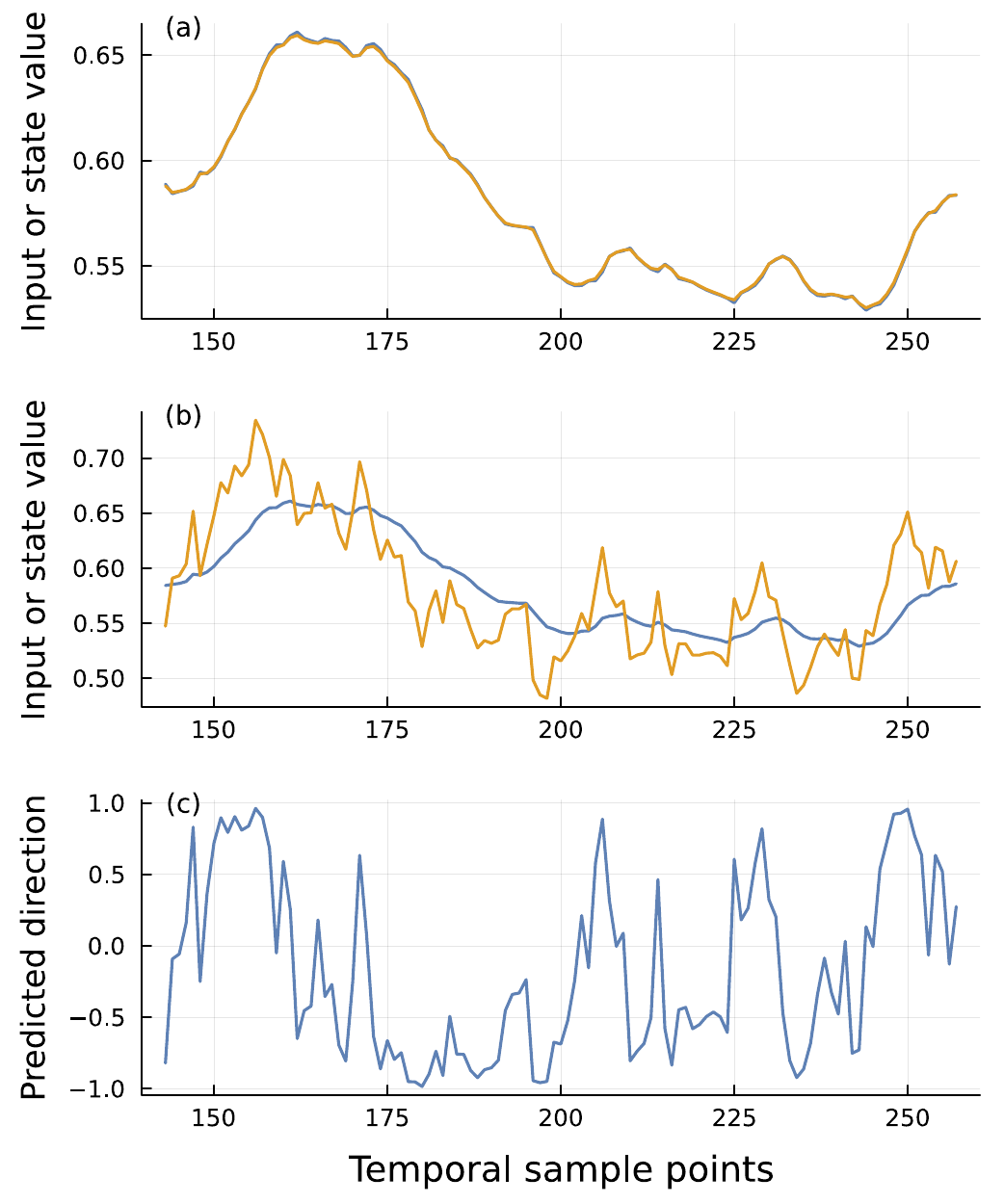}
\vskip0pt
\caption{Mechanism used by the machine learning encoder circuit to predict the direction of change in inputs. A sequence of inputs was generated as described in \Fig{enc_dyn}. These plots show a subset of the sample points. The circuit has two internal memory states, described in \Fig{encoder_net}. (a) The blue curve traces the prior input value. The gold curve shows one of the encoder circuit's internal memory states multiplied by a constant. (b) The other internal state rises and falls with the estimated momentum of the trend, shown in gold. I transformed the state value \reviset{to have the same mean as the input curve, which associated a positive internal momentum estimate (gold) with a location above the input (blue) curve and a negative momentum estimate with a location below the input curve. The blue curve is the same as in the upper panel.} (c) The predicted direction rises and falls with a function of the momentum. The sign provides the predicted direction. The magnitude describes an estimate for the confidence of the prediction. The actual circuit outputs predicted values $z$ on the interval $(0,1)$. Here, I show $2(z-0.5)$ so that positive and negative values correspond to positive and negative predicted changes.
}
\label{fig:enc_states}
\end{figure*}

\Figure{enc_states} describes the mechanism that the circuit uses to predict the direction of change. One internal memory state tracks the prior input. The other internal state is proportional to the momentum. Those states provide sufficient information to generate nearly optimal predictions about direction. The circuit also has information about its relative confidence in its predictions, given by the magnitude of the momentum estimate.

\subsection{Internal model}

Consider a trending numerical sequence, as in the previous examples. A circuit could learn to predict the direction of change in the next observation in two different ways.

First, the circuit could exploit patterns of correlation. If, for example, in the data from which the circuit learns, 8 typically follows the target sequence 2--4--5--9, then the circuit may learn to predict a negative change in response to the target. 
\begin{equation*}
	\text{learned correlation}\mskip40mu 2\mskip20mu 4\mskip20mu 5\mskip20mu 9\mskip20mu
		\fbox{8} \mskip12mu\implies\mskip12mu \text{\Large{$-$}}\\
\end{equation*}
Second, the circuit could learn an internal causal generative model for the sequence. For example, instead of focusing on the correlation between particular trigger sequences and the direction of future change, the circuit might learn the sort of causal model described in prior sections.
\begin{equation*}
	\text{causal model}\mskip40mu 2\mskip20mu 4\mskip20mu 5\mskip20mu 9\mskip20mu
		\fbox{X} \mskip12mu\implies\mskip14mu \text{\Large{+}}
\end{equation*}
Here, $X$ is predicted to be larger than $9$ because of the general tendency for trends to continue in their current direction. If a very simple circuit with limited memory can learn this underlying causal model, then it can predict the direction of change with nearly optimal accuracy.

Learning algorithms easily exploit correlations. If those correlations are just sampling artifacts in the original data, then the circuit will perform poorly in new situations. Sampling artifacts happen frequently in datasets that are small relative to their dimensionality, sometimes causing circuits to be fragile when focused on correlations.

Alternatively, if the learning dynamics can successfully reduce the dimensionality of the data in relation to the circuit's goal, then the circuit may be able to discover a relatively simple model that follows the contours of the underlying causal process. A good internal causal model performs well in response to novel input data.

\reviset{We do not know how often simple cellular circuits encode internal models of the environment rather than correlations between environmental states. An interesting literature develops the claim that internal models are particularly powerful solutions for prediction and control \autocite{bin22internal,gupta23theinternal}. Of course, given complete data on all possibles sequences and perfect predictions by correlational estimates, a correlational network converges to a perfect internal model. But, in practice, small computational networks may be able to calculate outcomes simply and quickly, whereas learning and computing highly accurate correlational predictions may be very costly or impossible.}

\subsection{A biochemical circuit for trend prediction}

This paper argues that machine learning and biological evolutionary learning may solve similar challenges with similar kinds of circuits. For example, the previous subsections showed how machine learning uses dimensional reduction to solve the problem of trend prediction. In this subsection, I use the insights from the machine learning solution to develop a hypothesis for how a simple biochemical circuit might solve the same problem.

In the machine learning solution, an internal model reduces the high dimensional input sequence to a single difference between the internal state estimates for the current input and a recent input. The sign of that difference predicts the direction of future change. The magnitude of that difference estimates the confidence in the prediction.

To build a biochemical circuit, we start with the machine learning solution's calculation of the difference between internal estimates for inputs at different time points. The following system of differential equations expresses a candidate biochemical circuit
\begin{align}\label{eq:cellTrend}
\begin{split}
  \dot{x}&=\Ga u - \Gb x\\
  \dot{y}&=\Gg\lr{\Ga u - \Gb y}\\
  \dot{z}&=\Gl + \eta(y - x) - \Gd z.
\end{split}
\end{align}
Overdots denote derivatives with respect to time, $t$. The variables $x,y,z,u$ are functions of time. Here, $u$ is a randomly changing input that follows the same process as described in the caption of \Fig{enc_dyn}.

For constant input, $u$, the molecular of abundances of $x$ and $y$ have the same equilibrium values, $\Ga/\Gb$. As the input, $u$, changes, $y$ changes more slowly than $x$ when $\Gg < 1$. Thus, $y-x$ tends to be negative when $u$ is increasing and positive when $u$ is decreasing. Therefore, the deviation of $z$ from its equilibrium, $\Gl/\Gd$, estimates the direction and magnitude of recent changes in $u$, providing a prediction for the direction of future change.

\begin{figure*}[t]
\centering
\vskip-1.5cm
\includegraphics[width=0.73\hsize]{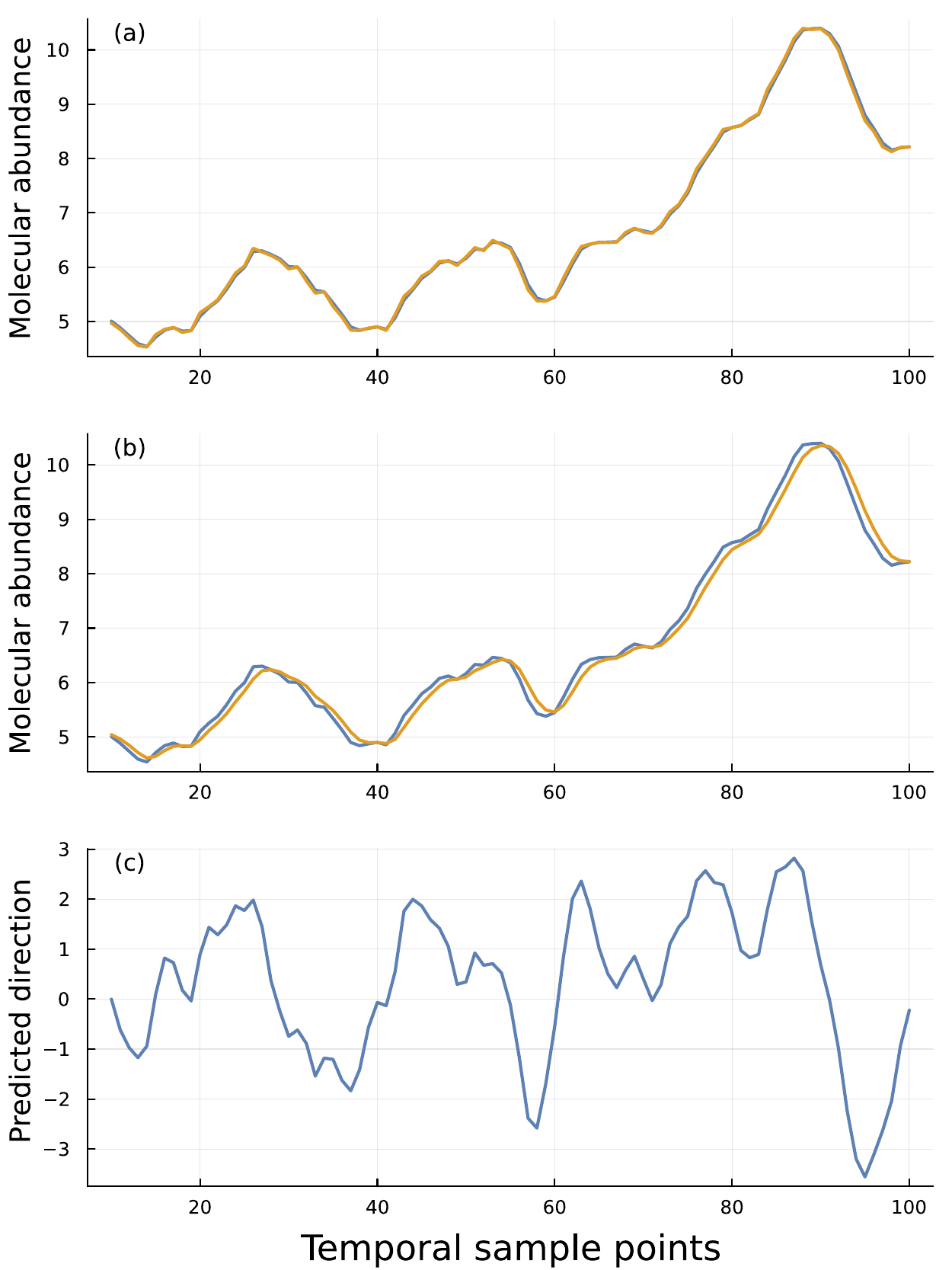}
\vskip0pt
\caption{Biochemical circuit prediction for the direction of change in a sequence of observations. Dynamics described by \Eq{cellTrend}. Parameters optimized by the method described below. (a) Molecular abundance of $x$ compared with the input sequence, $u$, in blue and gold, respectively. I multiplied $u$ by a constant to show the proportional match in the dynamics. The abundance of $x$ provides a good internal estimate of the recent input value. (b) Comparison of $x$ and $y$, in blue and gold, respectively. The abundance of $y$ responds more slowly to input changes. Thus, $y-x$ estimates the recent difference in input values. (c) The abundance of $z$ provides a prediction for the direction of change in the next sampled input. This panel shows the deviation of $z$ from its equilibrium multiplied by the parameter $\Gf$, as $\Gr=\Gf\large(z-\Gl/\Gd\large)$. The sign and magnitude of $\Gr$ predict the direction of change and the confidence in the prediction. In this example run, the sign of $\Gr$ predicted the sign of the next input change with $72.2\%$ accuracy compared with a true match between the sign of the difference in one increment and the sign of the difference in the next increment of $72.7\%$. I optimized the 6 parameters in \Eq{cellTrend} by minimizing the cross-entropy loss described in the caption of \Fig{encoder_net}. To minimize the loss function, I obtained values for the 6 parameters of \Eq{cellTrend} and for $\Gf$ by the Sophia algorithm with learning rate $0.005$ applied to 10,000 randomly generated input sequences \autocite{liu23sophia:}. Each input sequence was generated over 10 time units and sampled at $0.1$ time units, yielding 101 points. The times between points were interpolated linearly. To apply that input sequence to \Eq{cellTrend}, the time scaling for the inputs was multiplied by 3000 to yield a time span of 30,000 sampled at intervals of 300 time unit intervals to obtain the 101 sample points. In each run of the minimization routine and in these plots, I skipped the first 10 sample points.}

\label{fig:cellTrend}
\end{figure*}

Using the methods for input generation and discrete time point sampling in the caption of \Fig{enc_dyn}, I optimized the six parameters in \Eq{cellTrend}. \Figure{cellTrend} shows the results for a particular set of optimized parameters and a single realized sequence of the random input variable, $u$.

In this model, the internal states $x$ and $y$ approximately track a shorter and a longer exponential moving average of the inputs. The difference between those moving averages anticipates future trend.

By adjusting the memory window of those moving averages through the parameters $\Gb$ and $\Gg$, the system can modulate the timescales of trend estimation, future prediction, and the tradeoff between the overall accuracy and false signal frequency for trend reversals. Thus, this circuit provides a very simple and general way in which organisms can predict future environmental trends based on past environmental inputs \autocite{frank24a-biological}.

In retrospect, the structure of this anticipatory circuit seems obvious. However, I did not guess the solution until I observed how the machine learning circuit solved this problem.

For more complex problems, following the same method may often be helpful. First, use machine learning to search for how a circuit might solve the problem. Second, analyze how the machine learning circuit works. Does it embed an internal model? Third, construct and optimize a biological circuit that works in the same way. Finally, use the architecture of the biological model as a hypothesis for how actual biological circuits may be designed to solve the problem.

For the relatively small machine learning circuits that make good analogies for biochemical circuits, we can often reverse engineer the computational mechanism, as in this example.

Alternatively, we could start with a general system of differential equations that include a large pool of potential biochemical circuits. We then use machine learning methods directly on the system of equations to search for good biochemical circuit solutions to the specific challenge \autocite{hiscock19adapting,frank22optimization,frank23an-enhanced}.

\section{Dimensional reduction in biology}

Effective circuits reduce input dimension. Borges made this point clearly when, in his story ``Funes the Memorious,'' the title character recalled every detail of every moment of every day. Funes was incapacitated by his lack of reduction. The story's narrator concluded that ``To think is to forget a difference, to generalize, to abstract.'' \textcite{ogieblyn24thetrouble} summarized Funes incapacity:

\begin{quote}
Because Funes is incapable of forgetting any detail, he has problems understanding language. He cannot fathom why the word “dog” is used for so many distinct dogs, all of which vary in size, color, and form. He cannot even understand why the same dog is called by a single name—why, for instance, “the dog at three fourteen (seen from the side) should have the same name as the dog at three fifteen (seen from the front).”
\end{quote}

To learn is to exclude what does not matter. Dimensional reduction shaves to the invariant core. The issues concern: What is the best reduction for a given challenge? How is the reduction process embedded within the circuit? How do circuits learn to accomplish such reduction? Machine learning may become a helpful tool to answer these questions for biology because machine learning and biology face similar challenges \autocite{bengio13representation,chater08from,kemp08the-discovery,quiroga05invariant,tishby11information,levin24self-improvising}.

This section's examples illustrate the role of dimensional reduction in biological circuits. They also hint at the relations between the biological circuits discovered by natural selection and the engineered circuits discovered by machine learning \autocite{eckmann21dimensional,kanwisher23using,gamoran24detecting,mao24thetraining}.

\subsection{Metabolism, immunity, and development}

Bacterial cells switch metabolic pathways based on the internal concentrations of a few key molecules. These molecules serve as an internal representation, condensing the high-dimensional state of many nutrients into a simpler form used by the cell \autocite{kochanowski17few-regulatory,wu22cellular}.

Many biological circuits have an hourglass architecture. Diverse inputs cascade through the initial circuit layers toward a narrow middle processing layer. The diverse inputs may be thought of as the broad base of the hourglass, leading upward toward a narrow central waist. From the limited diversity in the narrow waist, an increasing number of different types flow upward toward the broad output layer at the top \autocite{csete04bow-ties,csete02reverse,doyle11architecture,frank23robustness,kaneko2024constructing}.

The narrow waist encodes a dimensionally reduced representation of the inputs, which the circuit decodes into a higher dimensional output. \textcite{friedlander15evolution} summarize examples of hourglass architectures in metabolism, immunity, and other biological pathways.

Metabolism starts with diverse nutrient inputs. Catabolic processing reduces those inputs to 12 common intermediates, such as pyruvate and fructose-6-phosphate. From those 12 intermediates, a cell makes most of its carbohydrates, nucleic acids, proteins, and other diverse biochemical forms \autocite{csete04bow-ties,zhao06hierarchical}.

Innate immune responses can be triggered by a wide diversity of molecular patterns associated with invasion. The immune system reduces those inputs to changes in a small number of toll-like receptors, interleukins, and a few other molecular types. That reduced immune encoding controls the broad response in gene expression and effector molecules that defend against attack \autocite{beutler04inferences,oda06acomprehensive,tang12pamp}.

Development often flows through an hourglass circuit \autocite{shook08morphogenic,irie14the-developmental,jordan23canalisation,frank23robustness}. Early developmental stages vary widely, forming the broad base of the hourglass. Intermediate stages reduce to fewer dimensions, the constrained waist. Later stages diversify to broad variations in adult form.

In the nematode \textit{Caenorhabditis elegans,} the developmental hourglass constrains how proteins explore the evolutionary landscape \autocite{ma23transcriptome}. Those proteins that affect early development evolve relatively rapidly when compared to related nematode species, the diverse hourglass base. Proteins that affect intermediate stages evolve relatively slowly, the dimensionally reduced hourglass waist. Proteins that affect late stages evolve relatively rapidly, the diverse hourglass top.

Doyle argued that this widely observed hourglass architecture arises inevitably in all complex, robust circuits \autocite{csete04bow-ties,csete02reverse,doyle11architecture,frank23robustness}. The dimensionally reduced narrow waist defines protocols that connect different layers in the processing flow.

For example, in metabolism, many different biochemical inputs are broken down into the same limited set of protocol-like core molecules. Those core molecules provide the basis for building new molecular types. The acquisition and breakdown of the building blocks create one layer, separate from the second layer that builds up functional forms. The digestion layer and buildup layer are connected by a dimensionally reduced common protocol core. The universality of metabolism arises from that reduced core.

\subsection{Neurobiology circuits}

We have the complete neural wiring diagram of \textit{C.\ elegans}. That connectome has an hourglass circuit architecture \autocite{sabrin20thehourglass,batta21aweighted}. Information from about 90 sensory neurons flows into a middle layer of approximately 80 recurrently connected interneurons. The third layer of about 120 motor neurons provides much of the network's response to inputs.

The middle layer includes 10--15 highly connected neurons that form the core hourglass waist, in which sensory inputs compress to a low dimensional encoding. Action follows from decoding the compressed information in the circuit waist, linking the broad sensory input and functional output layers through the narrow core protocol layer.

Rodent neural networks, vastly greater in size and connectivity, also have an hourglass circuit architecture \autocite{stuber23neurocircuits}. The sensory cortex takes in high dimensional information directly from all senses and from various intermediate processing ganglia. Those inputs are then compressed sequentially through a series of layers, including the limbic and frontal cortices, the hypothalamus, and the extended amygdala, leading finally to the greatest dimensional reduction in the core ventral tegmental area (VTA) dopamine layer.

The core VTA dopamine layer encodes the primary motivational protocols that transform the initial sensory inflow into a dopamine encoding for goal-directed behavior. The path from the core dopamine protocols to behavior flows through several decoding layers that expand dimensionality, including the ventral striatum, basal ganglia, and corticospinal motor circuits.

In machine learning, multilayer encoding and decoding subsystems are often beneficial for high-dimensional challenges. Each layer separately reduces or builds particular key dimensions in the encoder-decoder process that are difficult to learn or process simultaneously within single layers \autocite{goodfellow16deep}.

A study of human facial recognition compared how computational and biological networks parse facial features and differentiate individual faces \autocite{jiahui23modeling}. Both networks converged to essentially the same dimensional reductions for facial representation geometries and categorical face attributes. However, the networks differed in how they mapped the reduced encoding of facial features to the identification of individual faces.

The authors concluded that computational networks ``provide a good model for early cognitive and neural face processing of categorical attributes but are a poor model for individuation''.

I would rephrase that conclusion. A programmer makes many specific choices when building a computational model. The authors found that their particular computational choices for a deep convolutional neural network provided a good match for the neurobiological processing of facial categorical attributes. However, their particular computational implementation separated faces differently from how the neurobiological network separated faces.

It is likely that different implementations of computational models would differ in the degree to which they match the biological networks in categorical attributes and individuation. We might learn a lot about how both computational and biological networks function by studying how various computational implementations differ in their match to biology. It often happens that initial failures ultimately reveal the greatest insight.

For example, the ultimate performance metric in biology likely differs from the performance metric that a programmer might initially choose to make a computational model. \reviset{Specifically, biological design ultimately depends on reproductive success, whereas a programmer would likely use some measure for the success in parsing individual facial components or the success in discriminating between individuals. Reproductive success probably does not map linearly to a programmer's measures of facial parsing and discrimination.}

It would be interesting to know how different performance metrics alter machine learning circuits. That type of study might help to understand how different aspects of \reviset{dimensional reduction in facial or other input data} relate to function, providing testable hypotheses about biological circuits.

\section{Going forward}

We know how some biological circuits work. But we know relatively little about the full range of life's circuits. Reverse engineering is hard.

For some circuits in prokaryotes, we can list many of the molecules involved and their interaction partners. But the actual circuit designs with respect to information processing, dimensional reduction, and logic at a functional level remain open problems for many prokaryotic cellular processes. Eukaryotic cells and tissues are more complex.

Sometimes it's useful to take stock of what we don't know. Rapid advances in machine learning provide a basis for inventory, opening a way to explore how various circuit designs handle different challenges. In addition, we can easily alter constraints and performance metrics in machine learning to see how those factors change favored circuits. Such methods provide powerful tools for building a conceptual foundation and generating novel hypotheses.

Differences between machine learning and biology also provide interesting conceptual challenges. Are there intrinsic differences between natural selection and other learning algorithms in how they discover and design circuits? How do differing goals alter architecture?

Synthetic biology sets another challenge \autocite{braccini23recurrent}. How do we design biochemical circuits that achieve particular functions and performance goals? Machine learning provides a powerful set of tools. To use those tools effectively, we must figure out how to translate machine learning circuits into biochemical circuits or how to use machine learning tools directly to design synthetic biological networks \autocite{hiscock19adapting,frank22optimization,frank23an-enhanced,mishra23machine,nielsen16genetic,wu23machine}. 

%\section*{Acknowledgments}
%
%\noindent The Donald Bren Foundation, US Department of Defense grant W911NF2010227, and US National Science Foundation grant DEB-2325755 support my research.
%
%\section*{Data availability statement}
%
%Software to produce the figures is available on GitHub \autocite{frank24circuit-code}.

%\vfill\eject

\mybiblio	% uses main.bib by default, add other bibs as needed
\newpage

\begin{revisetenv}
\section*{Box 1: Reservoir computing}

Reservoir computing (RC) provides a model for how complex computations can emerge from seemingly simple network structures, mirroring the distributed processing observed in biological circuits.

The core of RC is the reservoir, a network of interconnected nodes, much like a network of neurons or a group of interacting molecules in a biochemical network of reactions. These nodes are linked by connections that allow information to flow between them. The connectivity is \textbf{recurrent}, meaning that information can flow in feedback loops. For example, node A can influence node B, which can influence node C, which in turn can loop back and influence node A again.

This circular flow of information, a continuous feedback loop, allows the system to maintain a memory of past signals and process information over time. Think of it as a dynamic echo chamber where incoming signals reverberate and interact, creating a complex internal representation of the input history.

Importantly, these internal connections are mostly randomly formed and static. They do not change during computation, and they are not improved or optimized as the system learns or evolves. Instead, they simply provide a passive vessel that stores a trace of past inputs. The inputs can be external environmental signals or internal signals arising from other components of the system.

Neuronal and gene regulatory networks also have recurrent connectivity, which plays an important role in processing temporal information. The random connectivity of reservoirs and the lack of change in connectivity over time suggest that precise, predetermined wiring is not necessary for memory or complex computation.

RC has several properties that may be important in biology. \textbf{Temporal memory} arises through recurrent connections, enabling the reservoir to maintain a fading memory of past inputs that determines the network's current state and response to new information. Similarly, a cell's current state reflects its history of environmental exposures.

\textbf{Distributed representation} occurs because information is not stored in a single location but instead is spread across the entire reservoir, often in a partially redundant manner. This distributed representation provides \textbf{robustness to noise and damage,} a valuable feature in biological networks.

The reservoir network is \textbf{tolerant to imperfect wiring} because the random connectivity and stasis mean that precise wiring is not a prerequisite for complex computation.

\textbf{Simplified learning} can happen because RC primarily involves adjusting how the reservoir's activity is interpreted rather than changing the connections within the reservoir itself. In particular, a system uses the stored information in the reservoir to adjust its response or predict a future input. The system can learn or evolve those responses while the reservoir network itself does not change.

RC typically has \textbf{emergent functional design} at a higher level relative to the partially random nature of lower level components. This architecture means that it may sometimes be difficult to infer function by starting at the lowest level and tracing upwards in a system.

\textbf{Overlapping circuits} may be common because the information stored in an RC network can be read out in different ways for different functions. An existing circuit may provide the basis for new functions by evolving novel responses that read out the circuit's stored information in a different way.

In summary, RC provides an interesting machine learning model for how evolution might have leveraged simple yet powerful design principles to create the complex information processing capabilities observed in living systems.

(I used various large language model systems to edit or generate parts of this text, including ChatGPT 4o, ChatGPT o1-preview, Claude 3.5 Sonnet (New), and Gemini Pro 1.5.)

\newpage

\section*{Box 2: Dimensional reduction}

Understanding how cells process complex environmental information into regulatory responses presents a fundamental challenge in biology. Machine learning, particularly the concept of dimensional reduction, offers powerful analogies for understanding how cellular networks accomplish this feat. Just as machine learning systems must distill vast amounts of complex data into meaningful patterns, cells must convert numerous environmental signals into specific regulatory decisions.

Dimensional reduction refers to the process of converting high-dimensional data into simpler, lower-dimensional representations while preserving essential information. Think of how a skilled artist can capture the essence of a complex scene in just a few brushstrokes. The artist has learned which features are essential to convey meaning and which can be omitted. Similarly, both machine learning systems and cells must learn to extract and represent the most important aspects of their environment while filtering out noise and redundancy.

In machine learning, \textbf{encoders} provide an elegant solution to this challenge. An encoder is like a skilled translator who must convey a story in fewer words while maintaining its essential meaning. The encoder learns to compress information by identifying patterns and relationships in the input data, creating a compact code that captures the key features. The power of this approach lies in how encoders discover efficient representations automatically through experience. They learn which combinations of features matter most for predicting outcomes.

For example, an encoder processing images of faces might learn to track key features like the distance between the eyes or the shape of the jaw, rather than storing every pixel. This compact representation contains the essential information needed to recognize or reconstruct the face. Similarly, cellular signaling networks may act as biological encoders, learning over evolutionary time to track key combinations of environmental signals rather than responding to each input independently.

The architecture of artificial encoders may provide clues for how cellular signaling cascades compress environmental information into key molecular signals, which regulatory networks then expand into broader cellular responses. Just as machine learning encoders must balance the trade-off between compression and preservation of important information, cellular networks face similar constraints in processing environmental signals efficiently with limited molecular resources.

In both machine learning and cellular systems, dimensional reduction through encoding offers several key advantages. It filters out noise and irrelevant variations, \textbf{making responses more robust.} It reduces the complexity of information processing, making it more energetically efficient. Perhaps most importantly, it \textbf{allows systems to generalize} from past experiences to new situations. Just as a machine learning model can recognize patterns in never-before-seen data, cellular networks can mount appropriate responses to novel combinations of environmental signals.

Viewing cellular regulatory networks through the lens of dimensional reduction provides valuable insights for biological research. It suggests new ways to analyze and model cellular decision-making, helps predict cellular responses to complex stimuli, and offers frameworks for understanding how cells maintain robust responses despite noisy inputs. This perspective also helps explain how relatively simple molecular networks can process complex environmental information and generate appropriate cellular responses.

(I used various large language model systems to edit or generate parts of this text, including ChatGPT 4o, ChatGPT o1-preview, Claude 3.5 Sonnet (New), and Gemini Pro 1.5.)

\end{revisetenv}
\newpage

%\afterpage{
%\clearpage
%}

%\afterpage{
%\clearpage
%}

%\afterpage{
%\clearpage
%}

% used cuted package strip env to force balancing of columns
%\ifmulticol\begin{strip}\hbox{\null}\end{strip}\hbox{\null}\fi

\end{document}

%% file: ml-def.tex
\newcommand*{\Ga}{\alpha}
\newcommand*{\Gb}{\beta}
\newcommand*{\Gd}{\delta}
\newcommand*{\GD}{\Delta}

\newcommand*{\Gg}{\gamma}

\newcommand*{\Gk}{\kappa}
\newcommand*{\Gl}{\lambda}

\newcommand*{\Gr}{\rho}
\newcommand*{\Gs}{\sigma}

\newcommand*{\Gf}{\phi}

\usepackage{bm}
\newcommand{\bmr}[1]{\bm{\mathrm{#1}}}

\DeclarePairedDelimiter\abs{\lvert}{\rvert}
\DeclarePairedDelimiter\norm{\lVert}{\rVert}
\DeclarePairedDelimiter\angb{\langle}{\rangle}
\DeclarePairedDelimiter\lrb{\lbrack}{\rbrack}
\DeclarePairedDelimiter\lr{\lparen}{\rparen}
\DeclarePairedDelimiter\lrbr{\lbrace}{\rbrace}

\makeatletter
\let\oldabs\abs \def\abs{\@ifstar{\oldabs}{\oldabs*}}
\let\oldnorm\norm \def\norm{\@ifstar{\oldnorm}{\oldnorm*}}
\let\oldangb\angb \def\angb{\@ifstar{\oldangb}{\oldangb*}}
\let\oldlrb\lrb \def\lrb{\@ifstar{\oldlrb}{\oldlrb*}}
\let\oldlr\lr \def\lr{\@ifstar{\oldlr}{\oldlr*}}
\let\oldlrbr\lrbr \def\lrbr{\@ifstar{\oldlrbr}{\oldlrbr*}}
\makeatother

\newcommand*{\dd}{\textrm{d}}
\newcommand*{\Eq}[1]{eqn~\ref{eq:#1}}

\newcommand*{\Figure}[1]{Figure~\ref{fig:#1}}
\newcommand*{\Fig}[1]{Fig.~\ref{fig:#1}}

\newcommand*{\boldrule}{\hrule height 1.2pt}
\newcommand*{\noterule}{\medskip\boldrule\medskip}	% for notes

%% file: ml.bbl
\begin{thebibliography}{10}
\expandafter\ifx\csname url\endcsname\relax
  \def\url#1{\texttt{#1}}\fi
\expandafter\ifx\csname urlprefix\endcsname\relax\def\urlprefix{URL }\fi
\providecommand{\bibinfo}[2]{#2}
\providecommand{\eprint}[2][]{\url{#2}}

\bibitem{wu22cellular}
\bibinfo{author}{Wu, C.} \emph{et~al.}
\newblock \bibinfo{title}{Cellular perception of growth rate and the
  mechanistic origin of bacterial growth law}.
\newblock \emph{\bibinfo{journal}{Proceedings of the National Academy of
  Sciences}} \textbf{\bibinfo{volume}{119}}, \bibinfo{pages}{e2201585119}
  (\bibinfo{year}{2022}).

\bibitem{goodfellow16deep}
\bibinfo{author}{Goodfellow, I.}, \bibinfo{author}{Bengio, Y.} \&
  \bibinfo{author}{Courville, A.}
\newblock \emph{\bibinfo{title}{Deep Learning}} (\bibinfo{publisher}{MIT
  Press}, \bibinfo{address}{Cambridge, MA}, \bibinfo{year}{2016}).
\newblock \urlprefix\url{http://www.deeplearningbook.org}.

\bibitem{kanwisher23using}
\bibinfo{author}{Kanwisher, N.}, \bibinfo{author}{Khosla, M.} \&
  \bibinfo{author}{Dobs, K.}
\newblock \bibinfo{title}{Using artificial neural networks to ask ‘why’
  questions of minds and brains}.
\newblock \emph{\bibinfo{journal}{Trends in Neurosciences}}
  \textbf{\bibinfo{volume}{46}}, \bibinfo{pages}{240--254}
  (\bibinfo{year}{2023}).

\bibitem{aldarondo24a-virtual}
\bibinfo{author}{Aldarondo, D.} \emph{et~al.}
\newblock \bibinfo{title}{A virtual rodent predicts the structure of neural
  activity across behaviours}.
\newblock \emph{\bibinfo{journal}{Nature}} \textbf{\bibinfo{volume}{632}},
  \bibinfo{pages}{594--602} (\bibinfo{year}{2024}).
\newblock \urlprefix\url{https://doi.org/10.1038/s41586-024-07633-4}.

\bibitem{mitchell23aischallenge}
\bibinfo{author}{Mitchell, M.}
\newblock \bibinfo{title}{{AI’s} challenge of understanding the world}.
\newblock \emph{\bibinfo{journal}{Science}} \textbf{\bibinfo{volume}{382}},
  \bibinfo{pages}{eadm8175} (\bibinfo{year}{2023}).

\bibitem{baltussen24chemical}
\bibinfo{author}{Baltussen, M.~G.}, \bibinfo{author}{de~Jong, T.~J.},
  \bibinfo{author}{Duez, Q.}, \bibinfo{author}{Robinson, W.~E.} \&
  \bibinfo{author}{Huck, W. T.~S.}
\newblock \bibinfo{title}{Chemical reservoir computation in a self-organizing
  reaction network}.
\newblock \emph{\bibinfo{journal}{Nature}} \textbf{\bibinfo{volume}{631}},
  \bibinfo{pages}{549--555} (\bibinfo{year}{2024}).
\newblock \urlprefix\url{https://doi.org/10.1038/s41586-024-07567-x}.

\bibitem{mcculloch43a-logical}
\bibinfo{author}{McCulloch, W.~S.} \& \bibinfo{author}{Pitts, W.}
\newblock \bibinfo{title}{A logical calculus of ideas immanent in nervous
  activity}.
\newblock \emph{\bibinfo{journal}{Bulletin of Mathematical Biophysics}}
  \textbf{\bibinfo{volume}{5}}, \bibinfo{pages}{115--133}
  (\bibinfo{year}{1943}).

\bibitem{hebb49theorganization}
\bibinfo{author}{Hebb, D.~O.}
\newblock \emph{\bibinfo{title}{The Organization of Behavior}}
  (\bibinfo{publisher}{Wiley}, \bibinfo{address}{New York},
  \bibinfo{year}{1949}).

\bibitem{rumelhart86parallel}
\bibinfo{author}{Rumelhart, D.~E.} \& \bibinfo{author}{McClelland, J.~L.}
\newblock \emph{\bibinfo{title}{Parallel {D}istributed {P}rocessing:
  {E}xplorations in the {M}icrostructure of {C}ognition, {V}olume 1:
  {F}oundations}} (\bibinfo{publisher}{MIT Press}, \bibinfo{address}{Cambridge,
  Massachusetts}, \bibinfo{year}{1986}).

\bibitem{jiahui23modeling}
\bibinfo{author}{Jiahui, G.} \emph{et~al.}
\newblock \bibinfo{title}{Modeling naturalistic face processing in humans with
  deep convolutional neural networks}.
\newblock \emph{\bibinfo{journal}{Proceedings of the National Academy of
  Sciences}} \textbf{\bibinfo{volume}{120}}, \bibinfo{pages}{e2304085120}
  (\bibinfo{year}{2023}).

\bibitem{maass02real-time}
\bibinfo{author}{Maass, W.}, \bibinfo{author}{Natschläger, T.} \&
  \bibinfo{author}{Markram, H.}
\newblock \bibinfo{title}{Real-time computing without stable states: A new
  framework for neural computation based on perturbations}.
\newblock \emph{\bibinfo{journal}{Neural Computation}}
  \textbf{\bibinfo{volume}{14}}, \bibinfo{pages}{2531--2560}
  (\bibinfo{year}{2002}).

\bibitem{jaeger07echo}
\bibinfo{author}{Jaeger, H.}
\newblock \bibinfo{title}{Echo state network}.
\newblock \emph{\bibinfo{journal}{Scholarpedia}} \textbf{\bibinfo{volume}{2}},
  \bibinfo{pages}{2330} (\bibinfo{year}{2007}).

\bibitem{gauthier21next}
\bibinfo{author}{Gauthier, D.~J.}, \bibinfo{author}{Bollt, E.},
  \bibinfo{author}{Griffith, A.} \& \bibinfo{author}{Barbosa, W.~A.}
\newblock \bibinfo{title}{Next generation reservoir computing}.
\newblock \emph{\bibinfo{journal}{Nature Communications}}
  \textbf{\bibinfo{volume}{12}}, \bibinfo{pages}{1--8} (\bibinfo{year}{2021}).

\bibitem{cucchi22hands-on}
\bibinfo{author}{Cucchi, M.}, \bibinfo{author}{Abreu, S.},
  \bibinfo{author}{Ciccone, G.}, \bibinfo{author}{Brunner, D.} \&
  \bibinfo{author}{Kleemann, H.}
\newblock \bibinfo{title}{Hands-on reservoir computing: a tutorial for
  practical implementation}.
\newblock \emph{\bibinfo{journal}{Neuromorphic Computing and Engineering}}
  \textbf{\bibinfo{volume}{2}}, \bibinfo{pages}{032002} (\bibinfo{year}{2022}).

\bibitem{damicelli22brain}
\bibinfo{author}{Damicelli, F.}, \bibinfo{author}{Hilgetag, C.~C.} \&
  \bibinfo{author}{Goulas, A.}
\newblock \bibinfo{title}{Brain connectivity meets reservoir computing}.
\newblock \emph{\bibinfo{journal}{PLoS Computational Biology}}
  \textbf{\bibinfo{volume}{18}}, \bibinfo{pages}{e1010639}
  (\bibinfo{year}{2022}).

\bibitem{goudarzi13dna-reservoir}
\bibinfo{author}{Goudarzi, A.}, \bibinfo{author}{Lakin, M.~R.} \&
  \bibinfo{author}{Stefanovic, D.}
\newblock \bibinfo{title}{Dna reservoir computing: A novel molecular computing
  approach}.
\newblock In \bibinfo{editor}{Soloveichik, D.} \& \bibinfo{editor}{Yurke, B.}
  (eds.) \emph{\bibinfo{booktitle}{DNA Computing and Molecular Programming}},
  \bibinfo{pages}{76--89} (\bibinfo{publisher}{Springer International
  Publishing}, \bibinfo{address}{Cham}, \bibinfo{year}{2013}).

\bibitem{yahiro18a-reservoir}
\bibinfo{author}{Yahiro, W.}, \bibinfo{author}{Aubert-Kato, N.} \&
  \bibinfo{author}{Hagiya, M.}
\newblock \bibinfo{title}{A reservoir computing approach for molecular
  computing}.
\newblock In \emph{\bibinfo{booktitle}{ALIFE 2018: The 2018 Conference on
  Artificial Life}}, \bibinfo{pages}{31--38} (\bibinfo{year}{2018}).

\bibitem{seoane19evolutionary}
\bibinfo{author}{Seoane, L.~F.}
\newblock \bibinfo{title}{Evolutionary aspects of reservoir computing}.
\newblock \emph{\bibinfo{journal}{Philosophical Transactions of the Royal
  Society B}} \textbf{\bibinfo{volume}{374}}, \bibinfo{pages}{20180377}
  (\bibinfo{year}{2019}).

\bibitem{czegel21novelty}
\bibinfo{author}{Czégel, D.}, \bibinfo{author}{Giaffar, H.},
  \bibinfo{author}{Csillag, M.}, \bibinfo{author}{Futó, B.} \&
  \bibinfo{author}{Szathmáry, E.}
\newblock \bibinfo{title}{Novelty and imitation within the brain: a darwinian
  neurodynamic approach to combinatorial problems}.
\newblock \emph{\bibinfo{journal}{Scientific Reports}}
  \textbf{\bibinfo{volume}{11}}, \bibinfo{pages}{12513} (\bibinfo{year}{2021}).

\bibitem{loeffler21modularity}
\bibinfo{author}{Loeffler, A.} \emph{et~al.}
\newblock \bibinfo{title}{Modularity and multitasking in neuro-memristive
  reservoir networks}.
\newblock \emph{\bibinfo{journal}{Neuromorphic Computing and Engineering}}
  \textbf{\bibinfo{volume}{1}}, \bibinfo{pages}{014003} (\bibinfo{year}{2021}).

\bibitem{sole22evolution}
\bibinfo{author}{Solé, R.} \& \bibinfo{author}{Seoane, L.~F.}
\newblock \bibinfo{title}{Evolution of brains and computers: the roads not
  taken}.
\newblock \emph{\bibinfo{journal}{Entropy}} \textbf{\bibinfo{volume}{24}},
  \bibinfo{pages}{665} (\bibinfo{year}{2022}).

\bibitem{frank23precise}
\bibinfo{author}{Frank, S.~A.}
\newblock \bibinfo{title}{Precise traits from sloppy components: perception and
  the origin of phenotypic response}.
\newblock \emph{\bibinfo{journal}{Entropy}} \textbf{\bibinfo{volume}{25}},
  \bibinfo{pages}{1162} (\bibinfo{year}{2023}).

\bibitem{sorrells15making}
\bibinfo{author}{Sorrells, T.~R.} \& \bibinfo{author}{Johnson, A.~D.}
\newblock \bibinfo{title}{Making sense of transcription networks}.
\newblock \emph{\bibinfo{journal}{Cell}} \textbf{\bibinfo{volume}{161}},
  \bibinfo{pages}{714--723} (\bibinfo{year}{2015}).

\bibitem{frank17puzzles5}
\bibinfo{author}{Frank, S.~A.}
\newblock \bibinfo{title}{Puzzles in modern biology. {V}. {W}hy are genomes
  overwired?}
\newblock \emph{\bibinfo{journal}{F1000Research}} \textbf{\bibinfo{volume}{6}},
  \bibinfo{pages}{924} (\bibinfo{year}{2017}).

\bibitem{chattopadhyay20data-driven}
\bibinfo{author}{Chattopadhyay, A.}, \bibinfo{author}{Hassanzadeh, P.} \&
  \bibinfo{author}{Subramanian, D.}
\newblock \bibinfo{title}{Data-driven predictions of a multiscale lorenz 96
  chaotic system using machine-learning methods: reservoir computing,
  artificial neural network, and long short-term memory network}.
\newblock \emph{\bibinfo{journal}{Nonlinear Processes in Geophysics}}
  \textbf{\bibinfo{volume}{27}}, \bibinfo{pages}{373--389}
  (\bibinfo{year}{2020}).
\newblock \urlprefix\url{https://npg.copernicus.org/articles/27/373/2020/}.

\bibitem{chepuri24hybridizing}
\bibinfo{author}{Chepuri, R.}, \bibinfo{author}{Amzalag, D.},
  \bibinfo{author}{Antonsen, T.~M.} \& \bibinfo{author}{Girvan, M.}
\newblock \bibinfo{title}{{Hybridizing traditional and next-generation
  reservoir computing to accurately and efficiently forecast dynamical
  systems}}.
\newblock \emph{\bibinfo{journal}{Chaos: An Interdisciplinary Journal of
  Nonlinear Science}} \textbf{\bibinfo{volume}{34}}, \bibinfo{pages}{063114}
  (\bibinfo{year}{2024}).
\newblock \urlprefix\url{https://doi.org/10.1063/5.0206232}.
\newblock
  \eprint{https://pubs.aip.org/aip/cha/article-pdf/doi/10.1063/5.0206232/19978096/063114\_1\_5.0206232.pdf}.

\bibitem{martinuzzi22reservoircomputing.jl:}
\bibinfo{author}{Martinuzzi, F.}, \bibinfo{author}{Rackauckas, C.},
  \bibinfo{author}{Abdelrehim, A.}, \bibinfo{author}{Mahecha, M.~D.} \&
  \bibinfo{author}{Mora, K.}
\newblock \bibinfo{title}{Reservoircomputing.jl: an efficient and modular
  library for reservoir computing models}.
\newblock \emph{\bibinfo{journal}{arXiv}} \textbf{\bibinfo{volume}{2204.05117}}
  (\bibinfo{year}{2022}).

\bibitem{ballarin24reservoir}
\bibinfo{author}{Ballarin, G.} \emph{et~al.}
\newblock \bibinfo{title}{Reservoir computing for macroeconomic forecasting
  with mixed-frequency data}.
\newblock \emph{\bibinfo{journal}{International Journal of Forecasting}}
  \textbf{\bibinfo{volume}{40}}, \bibinfo{pages}{1206--1237}
  (\bibinfo{year}{2024}).
\newblock
  \urlprefix\url{https://www.sciencedirect.com/science/article/pii/S0169207023001085}.

\bibitem{walleshauser22predicting}
\bibinfo{author}{Walleshauser, B.} \& \bibinfo{author}{Bollt, E.}
\newblock \bibinfo{title}{Predicting sea surface temperatures with coupled
  reservoir computers}.
\newblock \emph{\bibinfo{journal}{Nonlinear Processes in Geophysics}}
  \textbf{\bibinfo{volume}{29}}, \bibinfo{pages}{255--264}
  (\bibinfo{year}{2022}).
\newblock \urlprefix\url{https://npg.copernicus.org/articles/29/255/2022/}.

\bibitem{li20multi-reservoir}
\bibinfo{author}{Li, Q.}, \bibinfo{author}{Wu, Z.}, \bibinfo{author}{Ling, R.},
  \bibinfo{author}{Feng, L.} \& \bibinfo{author}{Liu, K.}
\newblock \bibinfo{title}{Multi-reservoir echo state computing for solar
  irradiance prediction: A fast yet efficient deep learning approach}.
\newblock \emph{\bibinfo{journal}{Applied Soft Computing}}
  \textbf{\bibinfo{volume}{95}}, \bibinfo{pages}{106481}
  (\bibinfo{year}{2020}).
\newblock
  \urlprefix\url{https://www.sciencedirect.com/science/article/pii/S1568494620304208}.

\bibitem{loeffler20topological}
\bibinfo{author}{Loeffler, A.} \emph{et~al.}
\newblock \bibinfo{title}{Topological properties of neuromorphic nanowire
  networks}.
\newblock \emph{\bibinfo{journal}{Frontiers in Neuroscience}}
  \textbf{\bibinfo{volume}{14}}, \bibinfo{pages}{184} (\bibinfo{year}{2020}).

\bibitem{munoz-gomez21constructive}
\bibinfo{author}{Muñoz-Gómez, S.~A.}, \bibinfo{author}{Bilolikar, G.},
  \bibinfo{author}{Wideman, J.~G.} \& \bibinfo{author}{Geiler-Samerotte, K.}
\newblock \bibinfo{title}{Constructive neutral evolution 20 years later}.
\newblock \emph{\bibinfo{journal}{Journal of Molecular Evolution}}
  \textbf{\bibinfo{volume}{89}}, \bibinfo{pages}{172--182}
  (\bibinfo{year}{2021}).

\bibitem{frank23robustnessb}
\bibinfo{author}{Frank, S.~A.}
\newblock \bibinfo{title}{Robustness and complexity}.
\newblock \emph{\bibinfo{journal}{Cell Systems}} \textbf{\bibinfo{volume}{14}},
  \bibinfo{pages}{1015--1020} (\bibinfo{year}{2023}).

\bibitem{lynch24evolutionary}
\bibinfo{author}{Lynch, M.~R.}
\newblock \emph{\bibinfo{title}{Evolutionary Cell Biology: The Origins of
  Cellular Architecture}} (\bibinfo{publisher}{Oxford University Press},
  \bibinfo{address}{New York}, \bibinfo{year}{2024}).

\bibitem{chen21proteome}
\bibinfo{author}{Chen, Y.} \emph{et~al.}
\newblock \bibinfo{title}{Proteome constraints reveal targets for improving
  microbial fitness in nutrient‐rich environments}.
\newblock \emph{\bibinfo{journal}{Molecular Systems Biology}}
  \textbf{\bibinfo{volume}{17}}, \bibinfo{pages}{e10093}
  (\bibinfo{year}{2021}).
\newblock
  \urlprefix\url{https://www.embopress.org/doi/abs/10.15252/msb.202010093}.
\newblock \eprint{https://www.embopress.org/doi/pdf/10.15252/msb.202010093}.

\bibitem{frank22microbial}
\bibinfo{author}{Frank, S.~A.}
\newblock \emph{\bibinfo{title}{{Microbial Life History: The Fundamental Forces
  of Biological Design}}} (\bibinfo{publisher}{Princeton University Press},
  \bibinfo{year}{2022}).

\bibitem{lynch07the-origins}
\bibinfo{author}{Lynch, M.}
\newblock \emph{\bibinfo{title}{The Origins of Genome Architecture}}
  (\bibinfo{publisher}{Sinauer Associates}, \bibinfo{address}{Sunderland, MA},
  \bibinfo{year}{2007}).

\bibitem{frank07maladaptation}
\bibinfo{author}{Frank, S.~A.}
\newblock \bibinfo{title}{Maladaptation and the paradox of robustness in
  evolution}.
\newblock \emph{\bibinfo{journal}{PLoS ONE}} \textbf{\bibinfo{volume}{2}},
  \bibinfo{pages}{e1021} (\bibinfo{year}{2007}).

\bibitem{lynch12evolutionary}
\bibinfo{author}{Lynch, M.}
\newblock \bibinfo{title}{Evolutionary layering and the limits to cellular
  perfection}.
\newblock \emph{\bibinfo{journal}{Proceedings of National Academy Sciences
  USA}} \textbf{\bibinfo{volume}{109}}, \bibinfo{pages}{18851--18856}
  (\bibinfo{year}{2012}).

\bibitem{gray10irremediable}
\bibinfo{author}{Gray, M.~W.}, \bibinfo{author}{Luke{\v{s}}, J.},
  \bibinfo{author}{Archibald, J.~M.}, \bibinfo{author}{Keeling, P.~J.} \&
  \bibinfo{author}{Doolittle, W.~F.}
\newblock \bibinfo{title}{Irremediable complexity?}
\newblock \emph{\bibinfo{journal}{Science}} \textbf{\bibinfo{volume}{330}},
  \bibinfo{pages}{920--921} (\bibinfo{year}{2010}).

\bibitem{daniels08sloppiness}
\bibinfo{author}{Daniels, B.~C.}, \bibinfo{author}{Chen, Y.-J.},
  \bibinfo{author}{Sethna, J.~P.}, \bibinfo{author}{Gutenkunst, R.~N.} \&
  \bibinfo{author}{Myers, C.~R.}
\newblock \bibinfo{title}{Sloppiness, robustness, and evolvability in systems
  biology}.
\newblock \emph{\bibinfo{journal}{Current Opinion in Biotechnology}}
  \textbf{\bibinfo{volume}{19}}, \bibinfo{pages}{389--395}
  (\bibinfo{year}{2008}).

\bibitem{gutenkunst07universally}
\bibinfo{author}{Gutenkunst, R.~N.} \emph{et~al.}
\newblock \bibinfo{title}{Universally sloppy parameter sensitivities in systems
  biology models}.
\newblock \emph{\bibinfo{journal}{PLoS Computational Biology}}
  \textbf{\bibinfo{volume}{3}}, \bibinfo{pages}{e189} (\bibinfo{year}{2007}).

\bibitem{rodan10minimum}
\bibinfo{author}{Rodan, A.} \& \bibinfo{author}{Tino, P.}
\newblock \bibinfo{title}{Minimum complexity echo state network}.
\newblock \emph{\bibinfo{journal}{IEEE Transactions on Neural Networks}}
  \textbf{\bibinfo{volume}{22}}, \bibinfo{pages}{131--144}
  (\bibinfo{year}{2010}).

\bibitem{rodan12simple}
\bibinfo{author}{Rodan, A.} \& \bibinfo{author}{Ti{\v{n}}o, P.}
\newblock \bibinfo{title}{Simple deterministically constructed cycle reservoirs
  with regular jumps}.
\newblock \emph{\bibinfo{journal}{Neural Computation}}
  \textbf{\bibinfo{volume}{24}}, \bibinfo{pages}{1822--1852}
  (\bibinfo{year}{2012}).

\bibitem{tripathi23minimal}
\bibinfo{author}{Tripathi, S.}, \bibinfo{author}{Kessler, D.~A.} \&
  \bibinfo{author}{Levine, H.}
\newblock \bibinfo{title}{Minimal frustration underlies the usefulness of
  incomplete regulatory network models in biology}.
\newblock \emph{\bibinfo{journal}{Proceedings of the National Academy of
  Sciences}} \textbf{\bibinfo{volume}{120}}, \bibinfo{pages}{e2216109120}
  (\bibinfo{year}{2023}).

\bibitem{ruach23thesynaptic}
\bibinfo{author}{Ruach, R.}, \bibinfo{author}{Ratner, N.},
  \bibinfo{author}{Emmons, S.~W.} \& \bibinfo{author}{Zaslaver, A.}
\newblock \bibinfo{title}{The synaptic organization in the caenorhabditis
  elegans neural network suggests significant local compartmentalized
  computations}.
\newblock \emph{\bibinfo{journal}{Proceedings of the National Academy of
  Sciences}} \textbf{\bibinfo{volume}{120}}, \bibinfo{pages}{e2201699120}
  (\bibinfo{year}{2023}).

\bibitem{ramezanian-panahi22generative}
\bibinfo{author}{Ramezanian-Panahi, M.} \emph{et~al.}
\newblock \bibinfo{title}{Generative models of brain dynamics}.
\newblock \emph{\bibinfo{journal}{Frontiers in Artificial Intelligence}}
  \textbf{\bibinfo{volume}{5}}, \bibinfo{pages}{807406} (\bibinfo{year}{2022}).

\bibitem{reddy20analysis}
\bibinfo{author}{Reddy, G.~T.} \emph{et~al.}
\newblock \bibinfo{title}{Analysis of dimensionality reduction techniques on
  big data}.
\newblock \emph{\bibinfo{journal}{IEEE Access}} \textbf{\bibinfo{volume}{8}},
  \bibinfo{pages}{54776--54788} (\bibinfo{year}{2020}).

\bibitem{sorzano14asurvey}
\bibinfo{author}{Sorzano, C. O.~S.}, \bibinfo{author}{Vargas, J.} \&
  \bibinfo{author}{Montano, A.~P.}
\newblock \bibinfo{title}{A survey of dimensionality reduction techniques}.
\newblock \emph{\bibinfo{journal}{arXiv preprint arXiv:1403.2877}}
  (\bibinfo{year}{2014}).

\bibitem{scholkopf21toward}
\bibinfo{author}{Schölkopf, B.} \emph{et~al.}
\newblock \bibinfo{title}{Toward causal representation learning}.
\newblock \emph{\bibinfo{journal}{Proceedings of the IEEE}}
  \textbf{\bibinfo{volume}{109}}, \bibinfo{pages}{612--634}
  (\bibinfo{year}{2021}).

\bibitem{bin22internal}
\bibinfo{author}{Bin, M.} \emph{et~al.}
\newblock \bibinfo{title}{Internal models in control, bioengineering, and
  neuroscience}.
\newblock \emph{\bibinfo{journal}{Annual Review of Control, Robotics, and
  Autonomous Systems}} \textbf{\bibinfo{volume}{5}}, \bibinfo{pages}{55--79}
  (\bibinfo{year}{2022}).

\bibitem{gupta23theinternal}
\bibinfo{author}{Gupta, A.} \& \bibinfo{author}{Khammash, M.}
\newblock \bibinfo{title}{The internal model principle for biomolecular control
  theory}.
\newblock \emph{\bibinfo{journal}{IEEE Open Journal of Control Systems}}
  \textbf{\bibinfo{volume}{2}}, \bibinfo{pages}{63--69} (\bibinfo{year}{2023}).

\bibitem{chen23auto-encoders}
\bibinfo{author}{Chen, S.} \& \bibinfo{author}{Guo, W.}
\newblock \bibinfo{title}{Auto-encoders in deep learning—a review with new
  perspectives}.
\newblock \emph{\bibinfo{journal}{Mathematics}} \textbf{\bibinfo{volume}{11}},
  \bibinfo{pages}{1777} (\bibinfo{year}{2023}).

\bibitem{mao24thetraining}
\bibinfo{author}{Mao, J.} \emph{et~al.}
\newblock \bibinfo{title}{The training process of many deep networks explores
  the same low-dimensional manifold}.
\newblock \emph{\bibinfo{journal}{Proceedings of the National Academy of
  Sciences}} \textbf{\bibinfo{volume}{121}}, \bibinfo{pages}{e2310002121}
  (\bibinfo{year}{2024}).

\bibitem{frank24a-biological}
\bibinfo{author}{Frank, S.~A.}
\newblock \bibinfo{title}{A biological circuit to anticipate trend}.
\newblock \emph{\bibinfo{journal}{Evolution Letters}} \bibinfo{pages}{qrae027}
  (\bibinfo{year}{2024}).

\bibitem{hiscock19adapting}
\bibinfo{author}{Hiscock, T.~W.}
\newblock \bibinfo{title}{Adapting machine-learning algorithms to design gene
  circuits}.
\newblock \emph{\bibinfo{journal}{BMC Bioinformatics}}
  \textbf{\bibinfo{volume}{20}}, \bibinfo{pages}{214} (\bibinfo{year}{2019}).

\bibitem{frank22optimization}
\bibinfo{author}{Frank, S.~A.}
\newblock \bibinfo{title}{Optimization of transcription factor genetic
  circuits}.
\newblock \emph{\bibinfo{journal}{Biology}} \textbf{\bibinfo{volume}{11}},
  \bibinfo{pages}{1294} (\bibinfo{year}{2022}).

\bibitem{frank23an-enhanced}
\bibinfo{author}{Frank, S.~A.}
\newblock \bibinfo{title}{An enhanced transcription factor repressilator that
  buffers stochasticity and entrains to an erratic external circadian signal}.
\newblock \emph{\bibinfo{journal}{Frontiers in Systems Biology}}
  \textbf{\bibinfo{volume}{3}}, \bibinfo{pages}{1276734}
  (\bibinfo{year}{2023}).

\bibitem{ogieblyn24thetrouble}
\bibinfo{author}{O'Gieblyn, M.}
\newblock \bibinfo{title}{The trouble with reality}.
\newblock \emph{\bibinfo{journal}{New York Review of Books}}
  \textbf{\bibinfo{volume}{March 21, 2024}} (\bibinfo{year}{2024}).

\bibitem{bengio13representation}
\bibinfo{author}{Bengio, Y.}, \bibinfo{author}{Courville, A.} \&
  \bibinfo{author}{Vincent, P.}
\newblock \bibinfo{title}{Representation learning: A review and new
  perspectives}.
\newblock \emph{\bibinfo{journal}{IEEE Transactions on Pattern Analysis and
  Machine Intelligence}} \textbf{\bibinfo{volume}{35}},
  \bibinfo{pages}{1798--1828} (\bibinfo{year}{2013}).

\bibitem{chater08from}
\bibinfo{author}{Chater, N.} \& \bibinfo{author}{Brown, G.~D.}
\newblock \bibinfo{title}{From universal laws of cognition to specific
  cognitive models}.
\newblock \emph{\bibinfo{journal}{Cognitive Science}}
  \textbf{\bibinfo{volume}{32}}, \bibinfo{pages}{36--67}
  (\bibinfo{year}{2008}).

\bibitem{kemp08the-discovery}
\bibinfo{author}{Kemp, C.} \& \bibinfo{author}{Tenenbaum, J.~B.}
\newblock \bibinfo{title}{The discovery of structural form}.
\newblock \emph{\bibinfo{journal}{Proceedings of the National Academy of
  Sciences}} \textbf{\bibinfo{volume}{105}}, \bibinfo{pages}{10687--10692}
  (\bibinfo{year}{2008}).

\bibitem{quiroga05invariant}
\bibinfo{author}{Quiroga, R.~Q.}, \bibinfo{author}{Reddy, L.},
  \bibinfo{author}{Kreiman, G.}, \bibinfo{author}{Koch, C.} \&
  \bibinfo{author}{Fried, I.}
\newblock \bibinfo{title}{Invariant visual representation by single neurons in
  the human brain}.
\newblock \emph{\bibinfo{journal}{Nature}} \textbf{\bibinfo{volume}{435}},
  \bibinfo{pages}{1102--1107} (\bibinfo{year}{2005}).

\bibitem{tishby11information}
\bibinfo{author}{Tishby, N.} \& \bibinfo{author}{Polani, D.}
\newblock \bibinfo{title}{Information theory of decisions and actions}.
\newblock In \bibinfo{editor}{Cutsuridis, V.}, \bibinfo{editor}{Hussain, A.} \&
  \bibinfo{editor}{Taylor, J.} (eds.)
  \emph{\bibinfo{booktitle}{{Perception-Action Cycle}}},
  \bibinfo{pages}{601--636} (\bibinfo{publisher}{Springer},
  \bibinfo{address}{New York}, \bibinfo{year}{2011}).

\bibitem{levin24self-improvising}
\bibinfo{author}{Levin, M.}
\newblock \bibinfo{title}{Self-improvising memory: a perspective on memories as
  agential, dynamically reinterpreting cognitive glue}.
\newblock \emph{\bibinfo{journal}{Entropy}} \textbf{\bibinfo{volume}{26}},
  \bibinfo{pages}{481} (\bibinfo{year}{2024}).
\newblock \urlprefix\url{https://www.mdpi.com/1099-4300/26/6/481}.

\bibitem{eckmann21dimensional}
\bibinfo{author}{Eckmann, J.-P.} \& \bibinfo{author}{Tlusty, T.}
\newblock \bibinfo{title}{Dimensional reduction in complex living systems:
  Where, why, and how}.
\newblock \emph{\bibinfo{journal}{BioEssays}} \textbf{\bibinfo{volume}{43}},
  \bibinfo{pages}{2100062} (\bibinfo{year}{2021}).

\bibitem{gamoran24detecting}
\bibinfo{author}{Gamoran, A.}, \bibinfo{author}{Lieberman, L.},
  \bibinfo{author}{Gilead, M.}, \bibinfo{author}{Dobbins, I.~G.} \&
  \bibinfo{author}{Sadeh, T.}
\newblock \bibinfo{title}{Detecting recollection: Human evaluators can
  successfully assess the veracity of others’ memories}.
\newblock \emph{\bibinfo{journal}{Proceedings of the National Academy of
  Sciences}} \textbf{\bibinfo{volume}{121}}, \bibinfo{pages}{e2310979121}
  (\bibinfo{year}{2024}).
\newblock \urlprefix\url{https://www.pnas.org/doi/abs/10.1073/pnas.2310979121}.
\newblock \eprint{https://www.pnas.org/doi/pdf/10.1073/pnas.2310979121}.

\bibitem{kochanowski17few-regulatory}
\bibinfo{author}{Kochanowski, K.} \emph{et~al.}
\newblock \bibinfo{title}{Few regulatory metabolites coordinate expression of
  central metabolic genes in \textit{{E}scherichia coli}}.
\newblock \emph{\bibinfo{journal}{Molecular Systems Biology}}
  \textbf{\bibinfo{volume}{13}}, \bibinfo{pages}{903} (\bibinfo{year}{2017}).

\bibitem{csete04bow-ties}
\bibinfo{author}{Csete, M.} \& \bibinfo{author}{Doyle, J.}
\newblock \bibinfo{title}{Bow ties, metabolism and disease}.
\newblock \emph{\bibinfo{journal}{TRENDS in Biotechnology}}
  \textbf{\bibinfo{volume}{22}}, \bibinfo{pages}{446--450}
  (\bibinfo{year}{2004}).

\bibitem{csete02reverse}
\bibinfo{author}{Csete, M.~E.} \& \bibinfo{author}{Doyle, J.~C.}
\newblock \bibinfo{title}{Reverse engineering of biological complexity}.
\newblock \emph{\bibinfo{journal}{Science}} \textbf{\bibinfo{volume}{295}},
  \bibinfo{pages}{1664--1669} (\bibinfo{year}{2002}).

\bibitem{doyle11architecture}
\bibinfo{author}{Doyle, J.~C.} \& \bibinfo{author}{Csete, M.}
\newblock \bibinfo{title}{Architecture, constraints, and behavior}.
\newblock \emph{\bibinfo{journal}{Proceedings of the National Academy of
  Sciences}} \textbf{\bibinfo{volume}{108}}, \bibinfo{pages}{15624--15630}
  (\bibinfo{year}{2011}).

\bibitem{frank23robustness}
\bibinfo{author}{Frank, S.~A.}
\newblock \bibinfo{title}{Robustness increases heritability: implications for
  familial disease}.
\newblock \emph{\bibinfo{journal}{Evolution}} \textbf{\bibinfo{volume}{77}},
  \bibinfo{pages}{655--659} (\bibinfo{year}{2023}).
\newblock \urlprefix\url{https://doi.org/10.1093/evolut/qpac026}.
\newblock \bibinfo{note}{Qpac026}.

\bibitem{kaneko2024constructing}
\bibinfo{author}{Kaneko, K.}
\newblock \bibinfo{title}{Constructing universal phenomenology for biological
  cellular systems: an idiosyncratic review on evolutionary dimensional
  reduction}.
\newblock \emph{\bibinfo{journal}{Journal of Statistical Mechanics: Theory and
  Experiment}} \textbf{\bibinfo{volume}{2024}}, \bibinfo{pages}{024002}
  (\bibinfo{year}{2024}).

\bibitem{friedlander15evolution}
\bibinfo{author}{Friedlander, T.}, \bibinfo{author}{Mayo, A.~E.},
  \bibinfo{author}{Tlusty, T.} \& \bibinfo{author}{Alon, U.}
\newblock \bibinfo{title}{Evolution of bow-tie architectures in biology}.
\newblock \emph{\bibinfo{journal}{PLoS Computational Biology}}
  \textbf{\bibinfo{volume}{11}}, \bibinfo{pages}{e1004055}
  (\bibinfo{year}{2015}).

\bibitem{zhao06hierarchical}
\bibinfo{author}{Zhao, J.}, \bibinfo{author}{Yu, H.}, \bibinfo{author}{Luo,
  J.-H.}, \bibinfo{author}{Cao, Z.-W.} \& \bibinfo{author}{Li, Y.-X.}
\newblock \bibinfo{title}{Hierarchical modularity of nested bow-ties in
  metabolic networks}.
\newblock \emph{\bibinfo{journal}{BMC Bioinformatics}}
  \textbf{\bibinfo{volume}{7}}, \bibinfo{pages}{1--16} (\bibinfo{year}{2006}).

\bibitem{beutler04inferences}
\bibinfo{author}{Beutler, B.}
\newblock \bibinfo{title}{Inferences, questions and possibilities in toll-like
  receptor signalling}.
\newblock \emph{\bibinfo{journal}{Nature}} \textbf{\bibinfo{volume}{430}},
  \bibinfo{pages}{257--263} (\bibinfo{year}{2004}).

\bibitem{oda06acomprehensive}
\bibinfo{author}{Oda, K.} \& \bibinfo{author}{Kitano, H.}
\newblock \bibinfo{title}{A comprehensive map of the toll-like receptor
  signaling network}.
\newblock \emph{\bibinfo{journal}{Molecular Systems Biology}}
  \textbf{\bibinfo{volume}{2}}, \bibinfo{pages}{2006--0015}
  (\bibinfo{year}{2006}).

\bibitem{tang12pamp}
\bibinfo{author}{Tang, D.}, \bibinfo{author}{Kang, R.}, \bibinfo{author}{Coyne,
  C.~B.}, \bibinfo{author}{Zeh, H.~J.} \& \bibinfo{author}{Lotze, M.~T.}
\newblock \bibinfo{title}{Pamps and damps: signals that spur autophagy and
  immunity}.
\newblock \emph{\bibinfo{journal}{Immunological Reviews}}
  \textbf{\bibinfo{volume}{249}}, \bibinfo{pages}{158--175}
  (\bibinfo{year}{2012}).

\bibitem{shook08morphogenic}
\bibinfo{author}{Shook, D.~R.} \& \bibinfo{author}{Keller, R.}
\newblock \bibinfo{title}{Morphogenic machines evolve more rapidly than the
  signals that pattern them: lessons from amphibians}.
\newblock \emph{\bibinfo{journal}{Journal of Experimental Zoology Part B:
  Molecular and Developmental Evolution}} \textbf{\bibinfo{volume}{310}},
  \bibinfo{pages}{111--135} (\bibinfo{year}{2008}).

\bibitem{irie14the-developmental}
\bibinfo{author}{Irie, N.} \& \bibinfo{author}{Kuratani, S.}
\newblock \bibinfo{title}{The developmental hourglass model: a predictor of the
  basic body plan?}
\newblock \emph{\bibinfo{journal}{Development}} \textbf{\bibinfo{volume}{141}},
  \bibinfo{pages}{4649--4655} (\bibinfo{year}{2014}).

\bibitem{jordan23canalisation}
\bibinfo{author}{Jordan, D.~J.} \& \bibinfo{author}{Miska, E.~A.}
\newblock \bibinfo{title}{Canalisation and plasticity on the developmental
  manifold of \textit{Caenorhabditis elegans}}.
\newblock \emph{\bibinfo{journal}{Molecular Systems Biology}}
  \textbf{\bibinfo{volume}{19}}, \bibinfo{pages}{e11835}
  (\bibinfo{year}{2023}).

\bibitem{ma23transcriptome}
\bibinfo{author}{Ma, F.} \& \bibinfo{author}{Zheng, C.}
\newblock \bibinfo{title}{Transcriptome age of individual cell types in
  \textit{{C}aenorhabditis elegans}}.
\newblock \emph{\bibinfo{journal}{Proceedings of the National Academy of
  Sciences}} \textbf{\bibinfo{volume}{120}}, \bibinfo{pages}{e2216351120}
  (\bibinfo{year}{2023}).

\bibitem{sabrin20thehourglass}
\bibinfo{author}{Sabrin, K.~M.}, \bibinfo{author}{Wei, Y.},
  \bibinfo{author}{van~den Heuvel, M.~P.} \& \bibinfo{author}{Dovrolis, C.}
\newblock \bibinfo{title}{The hourglass organization of the
  \textit{Caenorhabditis elegans} connectome}.
\newblock \emph{\bibinfo{journal}{PLoS Computational Biology}}
  \textbf{\bibinfo{volume}{16}}, \bibinfo{pages}{e1007526}
  (\bibinfo{year}{2020}).

\bibitem{batta21aweighted}
\bibinfo{author}{Batta, I.}, \bibinfo{author}{Yao, Q.},
  \bibinfo{author}{Sabrin, K.~M.} \& \bibinfo{author}{Dovrolis, C.}
\newblock \bibinfo{title}{A weighted network analysis framework for the
  hourglass effect—and its application in the \textit{C.\ elegans}
  connectome}.
\newblock \emph{\bibinfo{journal}{Plos ONE}} \textbf{\bibinfo{volume}{16}},
  \bibinfo{pages}{e0249846} (\bibinfo{year}{2021}).

\bibitem{stuber23neurocircuits}
\bibinfo{author}{Stuber, G.~D.}
\newblock \bibinfo{title}{Neurocircuits for motivation}.
\newblock \emph{\bibinfo{journal}{Science}} \textbf{\bibinfo{volume}{382}},
  \bibinfo{pages}{394--398} (\bibinfo{year}{2023}).

\bibitem{braccini23recurrent}
\bibinfo{author}{Braccini, M.}, \bibinfo{author}{Collinson, E.},
  \bibinfo{author}{Roli, A.}, \bibinfo{author}{Fellermann, H.} \&
  \bibinfo{author}{Stano, P.}
\newblock \bibinfo{title}{Recurrent neural networks in synthetic cells: a route
  to autonomous molecular agents?}
\newblock \emph{\bibinfo{journal}{Frontiers in Bioengineering and
  Biotechnology}} \textbf{\bibinfo{volume}{11}}, \bibinfo{pages}{1210334}
  (\bibinfo{year}{2023}).
\newblock
  \urlprefix\url{https://www.frontiersin.org/journals/bioengineering-and-biotechnology/articles/10.3389/fbioe.2023.1210334}.

\bibitem{mishra23machine}
\bibinfo{author}{Mishra, D.}, \bibinfo{author}{Ngam, P.~N.} \&
  \bibinfo{author}{Tyo, K.~E.}
\newblock \bibinfo{title}{Machine learning guided design and characterization
  of synthetic biology parts and circuits}.
\newblock \emph{\bibinfo{journal}{Metabolic Engineering}}
  \textbf{\bibinfo{volume}{75}}, \bibinfo{pages}{86--101}
  (\bibinfo{year}{2023}).

\bibitem{nielsen16genetic}
\bibinfo{author}{Nielsen, A.~A.} \emph{et~al.}
\newblock \bibinfo{title}{Genetic circuit design automation}.
\newblock \emph{\bibinfo{journal}{Science}} \textbf{\bibinfo{volume}{352}},
  \bibinfo{pages}{aac7341} (\bibinfo{year}{2016}).

\bibitem{wu23machine}
\bibinfo{author}{Wu, F.} \& \bibinfo{author}{Tan, C.}
\newblock \bibinfo{title}{Machine learning in synthetic biology: recent
  advances and future directions}.
\newblock \emph{\bibinfo{journal}{Nature Chemical Biology}}
  \textbf{\bibinfo{volume}{19}}, \bibinfo{pages}{635--646}
  (\bibinfo{year}{2023}).

\bibitem{frank24circuit-code}
\bibinfo{author}{Frank, S.~A.}
\newblock \bibinfo{title}{Circuit design in biology and machine learning. {I.
  Julia} software code} (\bibinfo{year}{2024}).
\newblock
  \urlprefix\url{https://github.com/evolbio/Circuits_01/releases/tag/Version_1.0.2}.

\bibitem{liu23sophia:}
\bibinfo{author}{Liu, H.}, \bibinfo{author}{Li, Z.}, \bibinfo{author}{Hall,
  D.}, \bibinfo{author}{Liang, P.} \& \bibinfo{author}{Ma, T.}
\newblock \bibinfo{title}{Sophia: A scalable stochastic second-order optimizer
  for language model pre-training}.
\newblock \emph{\bibinfo{journal}{arXiv:2305.14342}}  (\bibinfo{year}{2023}).

\end{thebibliography}
